\documentclass[a4paper,10pt,twocolumn]{article}
\pdfoutput=1

\usepackage{cite}
\usepackage{amsmath,amssymb,amsfonts}
\usepackage{algorithmic}
\usepackage{graphicx}
\usepackage{textcomp}
\def\BibTeX{{\rm B\kern-.05em{\sc i\kern-.025em b}\kern-.08em
		T\kern-.1667em\lower.7ex\hbox{E}\kern-.125emX}}

\usepackage[utf8]{inputenc}
\usepackage[english]{babel}
\usepackage[T1]{fontenc}
\usepackage[top=2cm, bottom=2cm, inner=2cm, outer=2cm]{geometry}

\usepackage{graphicx}
\usepackage{amssymb}
\usepackage{amsmath}
\usepackage{hyperref}
\usepackage{caption}
\usepackage{subcaption}
\usepackage{booktabs}
\usepackage{tikz}
\usetikzlibrary{quantikz}

\usepackage{verbatim}

\newtheorem{theorem}{Theorem}
\newtheorem{remark}{Remark}
\newtheorem{assumption}{Assumption}

\newtheorem{definition}{Definition}
\newtheorem{example}{Example}
\usepackage{setspace}
\usepackage{color}
\usepackage{url}
\usepackage{algorithm}
\usepackage{algorithmic}
\usepackage{bm}

\DeclareMathOperator*{\minimize}{minimize}
\DeclareMathOperator*{\argmin}{arg\,min}
\DeclareUnicodeCharacter{2212}{-}

\usepackage{cleveref}

\newcommand{\transp}{\intercal}

\newcommand{\revision}[1]{#1}
\newcommand{\rev}[1]{#1}
\newcommand{\admm}{M-ADMM-H}
\newcommand{\twoadmm}{$2$-ADMM-H}
\newcommand{\threeadmm}{$3$-ADMM-H}

\begin{document}
	\title{Multi-block ADMM Heuristics for Mixed-Binary Optimization on Classical and Quantum Computers}
	
	\author{Claudio Gambella$^1$ \and Andrea Simonetto$^1$}
	\date{%
		\small
		$^1$IBM Research Ireland, Mulhuddart, Dublin 15, Ireland\\%
	}
	
	
	\setlength{\parskip}{2mm} 
	\setlength{\parindent}{0pt}

	\maketitle

\begin{abstract}
	Solving combinatorial optimization problems on current noisy quantum
	 devices is currently being advocated for (and restricted to) binary polynomial optimization with equality constraints via quantum heuristic approaches. This is achieved using, e.g., the variational quantum eigensolver (VQE) and the quantum approximate optimization algorithm (QAOA). We present a decomposition-based approach to extend the applicability of current approaches to ``quadratic plus convex'' mixed binary optimization (MBO) problems, so as to solve a broad class of real-world optimization problems. In the MBO framework, we show that the alternating direction method of multipliers (ADMM) can split the MBO into a binary unconstrained problem (that can be solved with quantum algorithms), and continuous constrained convex subproblems (that can be solved cheaply with classical optimization solvers). The validity of the approach is then showcased by numerical results obtained on several optimization problems via simulations with VQE and QAOA on the quantum circuits implemented in Qiskit, an open-source quantum computing software development framework.
\end{abstract}

\maketitle

\section{Introduction}

Mixed-Binary Optimization (MBO) has been studied for decades in Mathematical Programming, because of the widespread range of applications in several domains \cite{cordeau2006branch, riera2006solving, gambella2019stochastic}, and the inherent difficulties posed by integer variables. \rev{MBO is known to be NP-hard in the general case.}
 The MBO class is very broad, and tailored exact or heuristic solution approaches have been devised in classical computation, depending on the nature and structure of the specific formulation \cite{floudas1995nonlinear, belotti2013mixed}. Recently, the advances in universal quantum computing~\cite{mcclean2016theory,moll2018quantum, farhi2014quantum, guerreschi2019qaoa} fostered efforts to understand whether this alternative computing paradigm could offer advantages (e.g., faster exact algorithms, more reliable heuristics) to solving combinatorial optimization problems \cite{zahedinejad2017combinatorial}. Research directed to apply the resulting algorithms to early generation of universal quantum computers has mainly focused on  quantum variational approaches \cite{mcclean2016theory}, which have been applied to chemistry \cite{peruzzo2014variational, kandala2017hardware}, machine learning \cite{romero2017quantum, farhi2018classification}, mathematical optimization \cite{moll2018quantum, fried2018qtorch, farhi2017quantum}. In broad terms, a variational approach works by choosing a parametrization of the space of quantum states that depends on a relatively small set of parameters, then using classical optimization routines to determine values of the parameters corresponding to a quantum state that maximizes or minimizes a given utility function. Typically, the utility function is given by a Hamiltonian encoding the total energy of the system, to be minimized. The variational theorem ensures that the expectation value of the Hamiltonian is greater than or equal to the minimum eigenvalue of the Hamiltonian. Such variational approaches can be applied for solving combinatorial optimization problems, provided that we can construct a Hamiltonian encoding the objective function of the optimization problem, see~\cite{barkoutsos2019improving, nannicini2019performance}. 
In the mathematical optimization context, research has been directed mainly to quadratic unconstrained binary optimization problems (QUBO):
\begin{align*}
	\minimize_{x} && c^{\transp} x + x^{\transp} Q x &&\\
	\text{subject to:}&& x \in \{0,1\}^n,&& 	\text{with $c \in \mathbb{R}^n, Q \in \mathbb{R}^{n \times n}$,}
\end{align*}
which can be transformed into an Ising model with Hamiltonian constituted as a summation of weighted tensor products of $Z$ Pauli operators. In case equality constraints $Ax=b$ are required to be modeled, a QUBO can still be devised by adding a quadratic penalization $\alpha \lVert Ax-b \rVert^2$ of the equality constraints to the objective function, as a soft-constraint in an Augmented Lagrangian fashion \cite{fiacco1990nonlinear, wang2009unified, nannicini2019performance}. 

A typical quantum variational approach, such as VQE~\cite{peruzzo2014variational} would involve two key steps in solving a QUBO, given its Ising Hamiltonian $H$ formulation. First, one would parametrize the quantum state via a small set of rotation parameters $\theta$: the state can then be expressed as $|\psi(\theta)\rangle = U(\theta)|0\rangle$, where $U(\theta)$ is the parametrized quantum circuit applied to the initial state $|0\rangle$. The variational approach would then aim at solving $\min_{\theta} \langle\psi(\theta)| H |\psi(\theta)\rangle$. Such optimization can be performed in a hybrid setting that uses a classical computer running an iterative algorithm to select $\theta$, and a quantum computer to compute information about $\langle\psi(\theta)| H |\psi(\theta)\rangle$ for given $\theta$ (e.g., its gradient). 

\rev{ 
In the MBO formulations, continuous variables and inequality constraints are typically both required to be modeled. Tackling a general MBO problem with quantum variational
approaches is at its early stages. One possibility is to introduce slack-based formulations, and consider the slacks as additional continuous parameters for the quantum QUBO solvers \cite{braine2019quantum}. While Grover searches have been applied for some continuous optimization problems \cite{PROTOPOPESCU20029}, and quantum annealing allows for tackling inequality constraints \cite{vyskovcil2019embedding}, the potential of quantum optimization algorithms for subclasses of MBO problems has not been investigated with a principled approach yet.
}

In this paper, we aim at extending the quantum optimization methodologies to be able to cope with MBOs on current quantum devices. As a matter of fact, we start in a bit more general context and we pose as an assumption that \rev{an approximate or noisy QUBO solver oracle is available to approximately solve QUBOs with some degree of sub-optimality; then, we ask ourselves where we can go from there. In particular, we explore ways to (heuristically) approximately solve certain classes of MBOs with the assumed noisy QUBO solver.} The aim of the paper is:
\begin{itemize}
\item To extend the quantum optimization methodologies to cope with MBOs;
\item To propose new heuristics to solve MBOs on quantum computers, having the potential to scale, in the future, to larger sizes than heuristics on classical computers;
\item To offer a glimpse on current research in combinatorial optimization in quantum computing, along with assumptions, challenges, and open problems. 
\end{itemize}

The proposed heuristics are based on the celebrated alternating direction method of multipliers (ADMM), see~\cite{boyd2011distributed,Eckstein1989,Bertsekas1997,Schizas2008,Glowinski1975,Gabay1976,He2011,Davis2014,Nishihara2015,Ghadimi2015,Deng2012,Giselsson2017}. Alternating Direction Method of Multipliers (ADMM) is an operator splitting algorithm that has a long history in convex optimization. ADMM is known to have residual, objective and dual variable convergence properties, provided that convexity assumptions are holding \cite{boyd2011distributed}. Recently, non-convex variants of ADMM have been developed (see~\cite{diamond2018general,wang2019global,sun2019two,Wu2019,Hong2016,Themelis2018,Takapoui2017} for non-convex/combinatorial theoretical results and ADMM-inspired heuristics). In the heuristic framework of \cite{diamond2018general} for problems with convex objective and decision variables from a non-convex set, ADMM relies on a (eventually approximate) projection on the non-convex feasibility set, and the ADMM iterates are improved via local search methods. Our method instead does not involve a projection step and makes use of the ADMM operator-splitting procedure to devise a decomposition for certain classes of MBOs into:
\begin{itemize}
	\item a QUBO subproblem to be solved by a QUBO \rev{(approximate) solver, e.g., on noisy quantum devices via quantum variational algorithms, such as VQE \cite{peruzzo2014variational}, QAOA \cite{farhi2014quantum}, or with Grover-search-based algorithms~\cite{Gilliam2019}, or quantum-based semidefinite programming (SDP) relaxations~\cite{Brandao2019}.}
	\item a convex constrained subproblem, which can be efficiently solved with classical optimization solvers \cite{boyd2004convex} . 
\end{itemize} 
Our method builds upon the recent results of \cite{wang2019global} and \cite{sun2019two}, which propose global convergence guarantees for non-convex and non-smooth optimization problems. For MBO, the convergence results of \cite{wang2019global} would not hold because of the requirement on the Lipschitz continuity of specific components of the objective function. However, \rev{\cite{Jiang2019,sun2019two}} observed that a third block can be added to the two-block decomposition of ADMM in order to gain convergence properties to stationary points. Our method also leverages the recent results of \cite{attouch2013convergence} on related tame problems to ensure that convergence is attained to a unique stationary point, by taking advantage of the semi-algebraic structure of the problem. \rev{Possible extensions to our current setting might be offered by \cite{Melo2017,Wang2019}, but not pursued here.}

The mathematical contribution of the paper is a multi-block ADMM heuristic (\admm) algorithm for MBO, for which we present:
\begin{itemize}
	\item a decomposition approach suitable for computation on current quantum devices;
	\item conditions for convergence, feasibility and optimality.
\end{itemize}

\rev{The quantum computing contribution of the paper is instead the analysis of the mentioned heuristic on quantum devices, and in particular:
\begin{itemize}
	\item computational results for classical and quantum implementations, including comments on the solution quality achieved;
	\item analysis of errors coming from an approximate solution of the underlying QUBO problem, which are typical in noisy (quantum) machines. 
\end{itemize}
}

Despite the quantum-oriented angle of the paper, our results applied also to classical computing, whenever a classical QUBO solver is available, or whenever the QUBO subproblems are easy or trivially solvable.    

The paper is organized as follows. \Cref{sec:alg-qc}~outlines current research effort in optimization in quantum computing, especially for QUBO solvers and the possible speed-up with respect to classical computing. \Cref{sec:alg-background}~reviews the ADMM proposed for convex optimization and introduces a two and three-block implementation of ADMM when binary variables are present. In \Cref{sec:alg-outline}, the two and three-block implementations are then specified for mixed-binary optimization problems and convergence properties are illustrated. To get a better picture of the proposed algorithm, \Cref{sec:simple-ex} discusses small-sized examples. The two illustrative MBO formulations in \Cref{sec:MBO} are solved with the proposed ADMM-based algorithms in \Cref{sec:comp-res}, discussing the results obtained on simulated quantum devices. Finally, conclusions are drawn in \Cref{sec:concl}.

\textbf{Notation.} Notation is whenever possible standard and borrowed from convex analysis~\cite{Rockafellar1970, boyd2004convex}. Vectors $x\in\mathbb{R}^n$, matrices $A \in \mathbb{R}^{n \times m}$, and sets $\mathcal{X} \subseteq \mathbb{R}^n$. For vectors and matrices $(\cdot)^\transp$ indicates the transpose operation. Functions $f: \mathbb{R}^n \to \mathbb{R}\cup \{+\infty\}$, whose codomain is the extended real line. 
A function is convex iff its epigraph is a convex set. A convex function is closed if its epigraph is closed, and it is proper if its effective domain is nonempty and it never attains $-\infty$. A function is lower-semicontinuous if and only if all of its sublevel sets $\{x\in \mathcal{X}:~f(x)\leq y\}$ are closed. A proper convex function is closed iff it is lower-semicontinuous. 
The indicator function of the set $\mathcal{X}$ is a function $\iota_{\mathcal{X}}:\mathbb{R}^n \to \mathbb{R}\cup \{+\infty\}$, for which $\iota_{\mathcal{X}}(x) = 0$ if $x \in \mathcal{X}$ and $+\infty$ otherwise. The indicator function of any closed set is lower-semicontinuous.

\section{Quantum computing for QUBOs} \label{sec:alg-qc}

\rev{ 
We briefly review here the current efforts in solving QUBOs via quantum/classical approaches. The aim of this section is to further expand on the state of the art, discuss what it is meant with potential quantum advantage of quantum solvers vs. classical solvers and why this is important for MBOs.} 

\rev{First of all, we look at current noisy quantum computers and quantum optimization algorithms, in particular VQE and QAOA. We do not touch upon quantum annealing, but the interested reader can find studies in the works of~\cite{Bian2014,Rosenberg2016,Karimi2017,Shaydulin2019} and references therein. VQE and QAOA are currently under major scrutiny and discussions on their performance, e.g.,~\cite{Brandao2018,Huang2019,Shaydulin2019a,guerreschi2019qaoa,nannicini2019performance,Zhou2019}, especially for the solution of unconstrained binary optimization problems, of which MaxCut is one important embodiment. On one hand, it seems that, in general, QAOA performance is still not well understood under all circumstances, and (even) some local classical algorithms could outperform it in some cases~\cite{Hastings2019}. On the other hand, some encouraging results in terms of performance advantage of QAOA with respect to the classical Goemans-Williamson limit have appeared~\cite{Crooks2018}. \revision{In addition, the combination of hyperparametrization and multi-start strategies has shown promising results in escaping local optima \cite{shaydulin2019multistart}}. These considerations, together with the fact that QAOA is not efficiently simulatable by classical computers, make QAOA an appealing algorithm to explore on noisy quantum machines.}

\rev{Second, looking ahead to less noisy quantum computers, quadratic speedup for Grover-based quantum QUBO solvers has been demonstrated in~\cite{Gilliam2019}, with respect to a classical unstructured search. While solving a large QUBO with an unstructured search is unrealistic for classical computers, we can envision employing such Grover-based quantum QUBO solvers for subproblems coming from classical branch-and-bound methodologies. This would generate a quantifiable speedup.} 

\rev{Third, looking further ahead to even less noisy quantum computers, quadratic speedup for quantum semidefinite programming (Q-SDP) relaxations arising in QUBO solving has been shown in~\cite{Brandao2019}. Given that currently SDP relaxations are one of the workhorses for QUBO solving, this new Q-SDP relaxations have the real potential of offering tangible speedup in solving QUBOs. } 

\rev{Given such wealth of results and studies, there is a significant interest to see if and how one can leverage quantum QUBO solvers to tackle more complicated constrained problems. This is what we are going to do next. 
}
\section{From the standard ADMM to a three-block structure}\label{sec:alg-background}

\subsection{Convex ADMM}

We start with some background on ADMM and the known results in the case of non-convex and combinatorial problems. Let $f: \mathbb{R}^n \to \mathbb{R}\cup \{+\infty\} $ and $h: \mathbb{R}^m \to \mathbb{R}\cup \{+\infty\}$ be closed convex proper functions, and let $A \in \mathbb{R}^{n \times p}, B \in \mathbb{R}^{m\times p}$ be given matrices. The prototypical problem we are interested in is of the form:
\begin{subequations}\label{eq:ADMM}
\begin{eqnarray}
\minimize_{x\in\mathbb{R}^n, y \in \mathbb{R}^m} && f(x) + h(y) \\
\mathrm{subject~to:} && A x + B y  = 0.
\end{eqnarray}
\end{subequations}
Then the ADMM is the following algorithm:
\begin{itemize}
\item Initialize the sequences $(x_k)_{k \in \mathbb{N}}$, $(y_k)_{k \in \mathbb{N}}$, $(\lambda_k)_{k \in \mathbb{N}}$ as $x_0\in \mathbb{R}^n$, $y_0\in \mathbb{R}^m$, $\lambda_0 \in \mathbb{R}^p$. Choose a penalty parameter $\varrho > 0$;
\item For $k = 1, 2, \dots$ do:
\begin{itemize}
\begin{subequations}\label{eq:ADMM:standard}
\item First block update:
\begin{eqnarray}
x_k = & \argmin_{x \in  \mathbb{R}^n} \, f(x) + \nonumber\\
& +  \lambda_{k-1}^{\transp}\,(A x + B y_{k-1}) + \nonumber\\
& + \frac{\varrho}{2}\|A x + B y_{k-1}\|^2_2;
\end{eqnarray}
\item Second block update:
\begin{eqnarray}
y_k = & \argmin_{y \in  \mathbb{R}^m} \, h(y) + \nonumber\\
 & +\lambda_{k-1}^{\transp}\,(A x_k + B y) + \nonumber\\
 & +  \frac{\varrho}{2}\|A x_k + B y\|^2_2;
\end{eqnarray}
\item Dual variable update:
\begin{equation}
\lambda_k = \lambda_{k-1} + \varrho (A x_{k} + B y_{k}).
\end{equation}
\end{subequations}
\end{itemize}
\end{itemize}

For the classical ADMM, we have various convergence and convergence rate results. For an ample classes of convex costs, ADMM converges for any $\varrho$, that is, starting from any $x_0, y_0, \lambda_0$, it generates a sequence for which we have
\begin{itemize}
\item Residual convergence: $A x_k + B y_k \to 0$ as $k \to \infty$, i.e., the iterates approach feasibility;
\item Objective convergence. $f(x_k) + h(y_k) \to p^*$ as $k \to \infty$, i.e., the objective function of the iterates approaches the optimal value;
\item Dual variable convergence. $\lambda_k \to \lambda^*$ as $k \to \infty$, where $\lambda^*$ is a dual optimal point.
\end{itemize}
See for instance~\cite{boyd2011distributed, Chen2017}, while for convergence rate analysis see for example~\cite{He2011, Giselsson2017}.

Non-convex results (when the cost functions are non-convex) are less ubiquitous in the literature and, in general, more restrictive in terms of assumptions. However, ADMM still behaves quite favourably in non-convex cases and attracts a considerable amount of attention from the research community.

\subsection{Mixed-binary ADMM}

In this paper, we start by modifying~\eqref{eq:ADMM} by considering that $x$ is now constrained to live in the non-convex set $\{0,1\}^n$, or equivalently that each of the component of the vector $x$, i.e., $x_{(i)}, i \in \{1, \dots, n\}$, is constrained as $x_{(i)}(1 - x_{(i)}) = 0$. We compactly write this as requiring $x \in \mathcal{X}$, where $\mathcal{X}$ represents the said binary set.

Let now $\iota_{\mathcal{X}}: \mathbb{R}^n \to \mathbb{R}\cup \{+\infty\}$ be the indicator function of the set $\mathcal{X}$, which is by construction closed and proper (but obviously non-convex), and consider the new function
$$
f^{\textrm{NC}}(x) = f(x) + \iota_{\mathcal{X}}(x).
$$
The function $f^{\textrm{NC}}(x)$ is non-convex by construction, yet one could still attempt at using the ADMM approach~\eqref{eq:ADMM:standard} with the new function $f^{\textrm{NC}}(x)$ in lieu of the ``old'' one $f(x)$, with the goal of solving the MBO:
\begin{subequations}\label{eq:ADMM:nc}
\begin{eqnarray}
\minimize_{x\in\mathcal{X}, y \in \mathbb{R}^m} && f(x) + h(y) \\
\mathrm{subject~to:} && A x + B y  = 0.
\end{eqnarray}
\end{subequations}
This is in general a heuristic. However, under some more restricting conditions the sequence generated by ADMM converges also in this case as follows. 

\makeatletter
\renewcommand*{\@opargbegintheorem}[3]{\trivlist
	\item[\hskip \labelsep{\bfseries #1\ #2}] \textbf{(#3)}\ \itshape}
\makeatother

\begin{theorem}[Convergence of mixed-binary ADMM~\cite{wang2019global}]\label{th:wang:2}
Consider the following assumptions:
\begin{itemize}
\item[\textbf{A1)}] (Coercivity) The objective function $f^{\textrm{NC}}(x) + h(y)$ is coercive over the set $Ax + By = 0$;
\item[\textbf{A2)}] (Feasibility) Im$(A) \subseteq $ Im$(B)$, where Im$(\cdot)$ returns the image of a matrix;
\item[\textbf{A3)}] (Lipschitz sub-minimization paths) There exists a positive constant $\bar{M}$, such that for any iterate counters $k_1$ and $k_2$, we have:
\begin{eqnarray}
\|x_{k_1} - x_{k_2}\| \leq \bar{M} \|A x_{k_1} - A x_{k_2}\|, \nonumber\\
 \|y_{k_1} - y_{k_2}\| \leq \bar{M} \|B y_{k_1} - B y_{k_2}\|;
\end{eqnarray}
\item[\textbf{A4)}] (Objective $f$-regularity) Function $f^{\textrm{NC}}(x)$ is lower semi-continuous;
\item[\textbf{A5)}] (Objective $h$-regularity) Function $h(y)$ is Lipschitz differentiable with constant $L_h$.
\end{itemize}
Define the augmented Lagrangian,
$$
\mathcal{L}_{\varrho}(x,y,\lambda) = f^{\textrm{NC}}(x) + h(y) + \lambda^{\transp}(Ax + By) + \frac{\varrho}{2}\|Ax + By\|^2_2. 
$$
Then, Binary ADMM converges subsequently for any sufficiently large $\varrho$, that is, starting from any $x_0, y_0, \lambda_0$, it generates a sequence that is bounded, has at least one limit point, and that each limit point $(x^*, y^*, \lambda^*)$ is a stationary point of $\mathcal{L}_{\varrho}$, namely, $0 \in \partial\mathcal{L}_{\varrho}(x^*, y^*, \lambda^*)$.

In addition, if $\mathcal{L}_{\varrho}$ is a Kurdyka-\L ojasiewicz (K\L) function \cite{Lojasiewicz1993,Bolte2007,Attouch2013}, then $(x_k, y_k, \lambda_k)$ converges globally to the unique limit point $(x^*, y^*, \lambda^*)$. 
\end{theorem}

Theorem~\ref{th:wang:2} is a special case of the more general Theorem~1 of~\cite{wang2019global} adapted to our problem setting (and where we have chosen to use a stronger version of A3) for sake of clarity and ease of implementation).  Functions satisfying the Kurdyka-\L ojasiewicz (K\L) property are for example semi-algebraic functions and locally strongly convex functions. We recall that a semi-algebraic function can be defined based on its graph as follows

\begin{definition}[\cite{Attouch2010}]
A subset of $\mathbb{R}^n$ is called semi-algebraic if it can be written as a finite union of sets of the form
$$
\{x\in\mathbb{R}^n : p_i(x)=0, q_i(x)<0, i=1,...,p\},
$$
where $p_i$, $q_i$ are real polynomial functions.

A function $f: \mathbb{R}^n \to \mathbb{R} \cup \{+\infty\}$ is semi-algebraic if its graph is a semi-algebraic subset of $\mathbb{R}^{n+1}$.
\end{definition} 

The following hold: (1) finite sums and products of semi-algebraic functions are semi-algebraic; (2) scalar products are semi-algebraic; (3) indicator functions of semi-algebraic sets are semi-algebraic; (4) generalized inverse of semi-algebraic mappings are semi-algebraic; (5) composition of semi-algebraic functions or mappings are semi-algebraic, see~\cite{Attouch2010}.

From this discussion, $\iota_{\mathcal{X}}(x)$ (besides being lower semi-continuous) is semi-algebraic, since its the indicator functions of semi-algebraic sets $\{x_{(i)}(1-x_{(i)}) = 0\}, \forall i$, and $f^{\textrm{NC}}(x)$ is semi-algebraic if $f(x)$ is semi-algebraic.

Theorem~\ref{th:wang:2} (or its broader version) is fairly tight, and counter-examples exists in which some of the assumptions are not verified and ADMM fails to converge. \rev{Relaxing some of the assumptions, for example A3), is a topic of current research, e.g., by leveraging the slightly different setting in~\cite{Melo2017}.}

To understand better the implications of~\ref{th:wang:2}, we consider a toy example, which verifies all the assumptions of the theorem. 

\begin{example}
Consider the problem: 
\begin{equation}
\min_{v \in \{0,1\}, w \in \mathbb{R}}\, -2 v + w^2, \, \textrm{subject to: } v = w.
\end{equation}

The unique optimal solution is $v^* = w^*= 1$. If we apply ADMM to it, as for Theorem~\ref{th:wang:2}, we can obtain convergence for sufficiently large $\varrho$ starting from any initial $v_0, w_0, \lambda_0$. For example, we can start with $v_0=1, w_0 = 1, \lambda_0 = 0$ with $\varrho = 100$. Then we can see that the ADMM algorithm converges to the solution $v =w = 0 =\lambda=  0$, which is a stationary point of the augmented Lagrangian $\mathcal{L}_{\varrho}(v,w,\lambda)$. If we start with a different starting point $v_0=0, w_0 = 0.5, \lambda_0 = 0$ with the same $\varrho$, then convergence is attained to the point $v =w = 1, \lambda=  2$, which is the optimal solution of the original problem, and another stationary point of the augmented Lagrangian.  
\end{example}

From the above example, one can understand the implications of convergence of ADMM in the non-convex setting, where one may converge to a feasible point, but not necessarily optimal for the problem. This is in general not a very unsatisfactory behaviour, especially in non-convex setting, where one is often concerned about finding ``good'' feasible points. 

\subsection{Mixed-binary three-block ADMM}

We move now to generalize the mixed-binary ADMM to three-block implementation. The reason behind the three blocks is that the assumptions in Theorem~\ref{th:wang:2} are restrictive for MBO problems and they would not be satisfied in general (as we see later).   

Consider the prototypical (mixed-binary) problem:
\begin{subequations}\label{eq:ADMM:3blocks}
\begin{eqnarray}
\minimize_{x\in\mathcal{X}, \bar{x}\in \mathbb{R}^l, y \in \mathbb{R}^m} && f_0(x) + f_1(\bar{x})+ h(y) \\
\mathrm{subject~to:} && A_0 x + A_1 \bar{x} + B y  = 0,
\end{eqnarray}
\end{subequations}
where we have introduced the functions $f_0: \mathbb{R}^n \to \mathbb{R}$, $f_1: \mathbb{R}^l \to \mathbb{R}$, the matrices $A_0 \in\mathbb{R}^{n \times p}$, $A_1 \in\mathbb{R}^{l \times p}$ and we have put ourselves already in the mixed-binary case. (For completeness, recall the definition of function $h: \mathbb{R}^m \to \mathbb{R}\cup \{+\infty\}$ as closed convex proper function, and matrix $B \in \mathbb{R}^{m\times p}$.) 

Then the three-block ADMM is the following algorithm: 
\begin{itemize}
\item Initialize the sequences $(x_k)_{k \in \mathbb{N}}$, $(\bar{x}_k)_{k \in \mathbb{N}}$, $(y_k)_{k \in \mathbb{N}}$, $(\lambda_k)_{k \in \mathbb{N}}$ as $x_0\in \mathbb{R}^n$, $\bar{x}_0\in \mathbb{R}^l$ $y_0\in \mathbb{R}^m$, $\lambda_0 \in \mathbb{R}^p$. Choose a penalty parameter $\varrho > 0$;
\item For $k = 1, 2, \dots$ do:
\begin{subequations}
\begin{itemize}
\item First block update:
\begin{eqnarray}
x_k = &  \argmin_{x \in  \mathbb{R}^n} \, f_0(x) + \iota_{\mathcal{X}}(x) + \nonumber\\
& + \lambda_{k-1}^{\transp}\,(A_0 x + A_1 \bar{x}_{k-1} + B y_{k-1}) + \nonumber\\
& +  \frac{\varrho}{2}\|A_0 x + A_1 \bar{x}_{k-1} + B y_{k-1}\|^2_2;
\end{eqnarray}
\item Second block update:
\begin{eqnarray}
\bar{x}_k = &  \argmin_{\bar{x} \in  \mathbb{R}^l} \, f_1(\bar{x}) + \nonumber\\
& + \lambda_{k-1}^{\transp}\,(A_0 x_k + A_1 \bar{x} + B y_{k-1}) + \nonumber\\
& +  \frac{\varrho}{2}\|A_0 x_k + A_1 \bar{x} + B y_{k-1}\|^2_2;
\end{eqnarray}
\item Third block update:
\begin{eqnarray}
y_k = & \argmin_{y \in  \mathbb{R}^m} \, h(y) + \nonumber\\
& + \lambda_{k-1}^{\transp}\,(A_0 x_k + A_1 \bar{x}_k + B y) + \nonumber\\ &+ \frac{\varrho}{2}\|A_0 x_k + A_1 \bar{x}_k + B y\|^2_2;
\end{eqnarray}
\item Dual variable update:
\begin{equation}
\lambda_k = \lambda_{k-1} + \varrho (A_0 x_{k} + A_1 \bar{x}_k + B y_{k}).
\end{equation}
\end{itemize}
\end{subequations}
\end{itemize}
This is in general a heuristic. However, under some more restricting conditions the sequence generated by ADMM converges also in this case as follows.

\begin{theorem}[Convergence of mixed-binary three-block ADMM~\cite{wang2019global}]\label{th:wang:3}
Consider the following assumptions:
\begin{itemize}
\item[\textbf{A1)}] (Coercivity) The objective function $f_0^{\textrm{NC}}(x) + f_1(\bar{x}) + h(y)$ is coercive over the set $A_0 x + A_1 \bar{x} + By = 0$, where we have defined $f_0^{\textrm{NC}}(x):= f_0(x) + \iota_{\mathcal{X}}(x)$;
\item[\textbf{A2)}] (Feasibility) Im$(A) \subseteq $ Im$(B)$, where $A = [A_0, A_1]$;
\item[\textbf{A3)}] (Lipschitz sub-minimization paths) There exists a positive constant $\bar{M}$, such that for any iterate counters $k_1$ and $k_2$, we have:
\begin{eqnarray}
\|x_{k_1} - x_{k_2}\| \leq \bar{M} \|A_0 x_{k_1} - A_0 x_{k_2}\|, \nonumber\\
 \|\bar{x}_{k_1} - \bar{x}_{k_2}\| \leq \bar{M} \|A_1 \bar{x}_{k_1} - A_1 \bar{x}_{k_2}\|, \nonumber\\
  \|y_{k_1} - y_{k_2}\| \leq \bar{M} \|B y_{k_1} - B y_{k_2}\|;
\end{eqnarray}
\item[\textbf{A4)}] (Objective $f$-regularity) Function $f_0^{\textrm{NC}}(x)$ is lower semi-continuous and $f_1(\bar{x})$ is restricted prox-regular; 
\item[\textbf{A5)}] (Objective $h$-regularity) Function $h(y)$ is Lipschitz differentiable with constant $L_h$.
\end{itemize}
Define the augmented Lagrangian,
\begin{align}
\mathcal{L}_{\varrho}(x,\bar{x},y,\lambda) & = f_0^{\textrm{NC}}(x) + f_1(\bar{x}) + h(y) + \nonumber\\
&  +\lambda^{\transp}(A_0 x + A_1 \bar{x} + By ) + \nonumber\\
& + \frac{\varrho}{2}\|A_0 x + A_1 \bar{x} + By \|^2_2. 
\end{align}
Then, Mixed-binary three-block ADMM converges subsequently for any sufficiently large $\varrho$, that is, starting from any $x_0, \bar{x}_0, y_0, \lambda_0$, it generates a sequence that is bounded, has at least one limit point, and that each limit point $(x^*,\bar{x}^*, y^*, \lambda^*)$ is a stationary point of $\mathcal{L}_{\varrho}$, namely, $0 \in \partial\mathcal{L}_{\varrho}(x^*, \bar{x}^*, y^*, \lambda^*)$.

In addition, if $\mathcal{L}_{\varrho}$ is a Kurdyka-\L ojasiewicz (K\L) function \cite{Lojasiewicz1993,Bolte2007,Attouch2013}, then $(x_k, \bar{x}_k, y_k, \lambda_k)$ converges globally to the unique limit point $(x^*, \bar{x}^*, y^*, \lambda^*)$.
\end{theorem}

Theorem~\ref{th:wang:3} is a special case of the more general Theorem~1 of~\cite{wang2019global} adapted to our problem setting (and where we have chosen to use a stronger version of A3) for sake of clarity and ease of implementation).  Functions satisfying the restricted prox-regularity assumptions are for example convex functions, including indicator functions of convex sets (which will be the ones that we will use in the sequel). 

What is now fundamental in the three-block ADMM is that we can restrict variable $x$ to be binary, and shift all the other constraints on $\bar{x}$ (any restricted prox-regular constraints, e.g., linear inequalities). This without affecting the variable $y$, which stays unconstrained, and whose function $h(y)$ needs to be smooth (so one cannot add an indicator function to represent additional constraint there). This ``trick'' was first explored in~\cite{sun2019two} in the context of distributed computations and discussed in the following example. 

\begin{example}\label{ex:2}
Consider the problem: 
\begin{equation}
\min_{v \in \{0,1\}, w \geq 1/2}\, -2 v + w^2, \, \textrm{subject to: } v = w.
\end{equation}

This problem does not satisfy the assumptions of Theorem~\ref{th:wang:2}, since $y$ is now constrained (although ADMM here is nonetheless converging in practice). But a possible way to overcome this (without adding constraints on the binary variable $v$), is to use the relaxed problem
\begin{equation}
\min_{v \in \{0,1\}, \bar{v} \geq 1/2, w \in \mathbb{R}}\, -2 v + \bar{v}^2 + \frac{\beta}{2} w^2, \, \textrm{subject to: } v = \bar{v} +  w,
\end{equation}
for a large $\beta< \varrho$. 

Starting $v_0 = \bar{v}_0 = w_0 = \lambda_0 = 0$ with $\beta = 1000$ and $\rho = 1001$, we obtain a sequence converging to $v = 1, \bar{v}=0.998, w= 0.002, \lambda = 1.996$, which is close to the optimal solution of the original problem. 

In~\cite{sun2019two}, a proper dualization of the constraint $w = 0$ is imposed, but the convergence of the then two-level approach has more restricting assumptions that the ones that we consider here, \rev{in particular $\bar{v}$ needs to be constrained in an hypercube~\cite[As. 4.2]{sun2019two}.}

\end{example}

\section{Two and three-block ADMM algorithms for MBO}\label{sec:alg-outline}

%
%

\subsection{From MBOs to two-block ADMM}

We are now ready to tackle MBOs. In this paper, we will consider the following reference problem $(P)$:
\begin{subequations}\label{eq:MBO}
\begin{eqnarray}
\minimize_{x \in \mathcal{X},u\in\mathcal{U} \subseteq \mathbb{R}^l } \ &&  q(x) + \varphi(u)  \\
\mathrm{subject~to:} && G x = b, \quad  g(x) \leq 0 \\
&&  \ell(x, u) \leq 0,
\end{eqnarray}
\end{subequations}
with the corresponding functional assumptions.

\begin{assumption}[Functional assumptions]\label{as:1} The following assumptions hold:
\begin{itemize}
	\item Function $q: \mathbb{R}^n \to \mathbb{R}$ is quadratic, i.e., $q(x) = x^{\transp} Q x + a^{\transp} x$ for a given symmetric squared matrix $Q \in \mathbb{R}^n \times \mathbb{R}^n, Q = Q^{\transp}$, and vector $a \in \mathbb{R}^n$;
	\item The set $\mathcal{X} = \{0,1\}^n = \{x_{(i)} (1-x_{(i)}) = 0, \forall i\}$ enforces the binary constraints;
	\item Matrix $G\in\mathbb{R}^n \times \mathbb{R}^{n'}$, vector $b \in \mathbb{R}^{n'}$, and function $g: \mathbb{R}^n \to \mathbb{R}$ is convex;
	\item Function $\varphi: \mathbb{R}^l \to \mathbb{R}$ is convex and $\mathcal{U}$ is a convex set;
	\item Function $\ell: \mathbb{R}^n\times  \mathbb{R}^l \to \mathbb{R}$ is \emph{jointly} convex in $x, u$.
\end{itemize}
\end{assumption}

Problem $(P)$ with the required functional assumptions can still capture many relevant problems in mathematical programming, such as vehicle routing \cite{toth2014vehicle, cordeau2006branch, bianchessi2014distance} and facility location \cite{campbell1994integer}. Formulations for bin packing and knapsack problems will be discussed in \Cref{sec:MBO}. 

In order to put Problem $(P)$ in the ADMM standard form, we need to write Problem $(P)$ as problem~\eqref{eq:ADMM:nc}. First, in this paper, following mainstream quantum practice (see \cite{nannicini2019performance, braine2019quantum}) and because we need to retrieve a QUBO, we soft-constrain the equality constraint (whenever present) as an augmented term in the cost function. Then, we introduce the new variable $z \in \mathbb{R}^n$ and Problem $(P)$ can be written as the soft-constrained problem $(P'):$ 
\begin{subequations}\label{eq:MBO:split}
\begin{eqnarray}
\hspace*{-1cm}\minimize_{x \in \mathcal{X},z\in \mathbb{R}^n, u\in\mathcal{U} \subseteq \mathbb{R}^l } \ &&  q(x) + \frac{c}{2}\|Gx - b\|^2_2 +\varphi(u)  \\
\mathrm{subject~to:~} && g(z) \leq 0, \quad  \ell(z, u) \leq 0, \\
&& x = z,
\end{eqnarray}
\end{subequations}
for a large positive constant $c>0$. Problem~$(P')$ is a soft-constrained version of Problem $(P)$ (it would be equivalent if $G = 0$ and $b = 0$): it is however a convenient splitting of binary and continuous variables.

Now, call $\bar{x} = [z^{\transp}, u^{\transp}]^{\transp}$, $m = n + l$, define $f_0(x) := q (x) + \frac{c}{2}\|Gx - b\|^2_2$, $f_1(\bar{x}) = \varphi(u) + \iota_{\bar{\mathcal{X}}}(\bar{x})$, where the set $\bar{\mathcal{X}}: = \{(z\in\mathbb{R}^n,u\in\mathcal{U})| g(z) \leq 0, \quad  \ell(z, u) \leq 0\}$. Then $(P')$ reads as problem $(P'')$:
\begin{subequations}\label{eq:MBO:final}
\begin{eqnarray}
 \minimize_{x \in \mathcal{X},\bar{x} \in \mathbb{R}^m } \ &&  f_0(x) + f_1(\bar{x})  \\
\mathrm{subject~to:} && A_0 x + A_1 \bar{x} = 0,
\end{eqnarray}
\end{subequations}
where $A_0 = I_n$ and $A_1 = [-I_n, 0_{l\times l}]$. 

A first possible strategy to use ADMM on $(P'')$ is summarized in Algorithm~\ref{algo:twoblock}, in a two-block implementation (\twoadmm). As we discussed in Section~\ref{sec:alg-background} and Example~\ref{ex:2}, this strategy is in general a heuristic, since the variable $\bar{x}$ is constrained, however in some cases Algorithm~\ref{algo:twoblock} can deliver good solutions (as we will explore). In order to keep track of the solution quality during the iterations, we compute a merit value associated with each iterate $x_k$. Let $\zeta_k = \max (g(x_k), 0) + \max(l(x_k, \bar{x}_k), 0)$ be the violation of the constraints on decision variable $x$ in Problem $(P)$ at iteration $k$, and $\mu$ be a penalization for $\zeta_k$. Then, the merit value $\eta_k$ of $x_k$ is a linear combination $ q(x_k)+\phi(\bar{x}_k) + \mu \zeta_k$  of the constraint violation and solution cost in problem $(P)$. Iterates with high merit value are both not likely to be of optimal value and close to feasibility for Problem $(P)$, hence the minimum merit value solution is returned by Algorithm~\ref{algo:twoblock}.

	\begin{algorithm}[H]
		\caption{\twoadmm~mixed-binary heuristic}\label{algo:twoblock} 
		\begin{algorithmic}[1] 
			\small{\REQUIRE Initial choice of $x_0, y_0, \lambda_0$. Choice of $\varrho, c, \mu>0$, tolerance $\epsilon>0$, and maximum number of iterations $K_{\text{max}}$.
				\WHILE {$k<K_{\text{max}}$ \AND $\|A_0 x^k+					A_1 \bar{x}^k\| > \epsilon,$}
				\STATE First block update (QUBO):
				\begin{multline*}
				x_k = \argmin_{x \in \{0,1\}^n} \,\, q (x) + \frac{c}{2}\|Gx - b\|^2_2 + \lambda_{k-1}^{\transp} A_0 x \\+ \frac{\varrho}{2}\|A_0 x + A_1 \bar{x}_{k-1} \|^2
				\end{multline*}
				\STATE Second block update (Convex):
				$$
				\bar{x}_k = \argmin_{\bar{x} \in \mathbb{R}^m } \,\, f_1(\bar{x}) + \lambda_{k-1}^{\transp} A_1 \bar{x} + \frac{\varrho}{2}\|A_0 x_{k} + A_1 \bar{x} \|^2
				$$
				\STATE Dual variable update:
				$$
				\lambda_k = \lambda_{k-1} + \varrho(A_0 x_{k} + A_1 \bar{x}_k)
				$$
				\STATE Compute merit value:
				\begin{multline*}
				\eta_k = q(x_k)+\phi(\bar{x}_k) \\+ \mu (\max (g(x_k), 0) + \max(l(x_k, \bar{x}_k), 0))
				\end{multline*}
				\ENDWHILE
						 \RETURN $x_{{k}^*}, \bar{x}_{{k}^*}, y_{{k}^*},$ with ${{k}^*} = \min_k \eta_k.$}
	\end{algorithmic}\end{algorithm}
	%

The strength of Algorithm~\ref{algo:twoblock} is that the original MBO is now split into a QUBO (that can be solved on the QUBO oracle, or on quantum devices) and a convex problem, that can be solved with off-the-shelf solvers, such as SPDT3 \cite{toh1999sdpt3} and MOSEK \cite{andersen2000mosek}.  

\begin{remark}
In~\cite{diamond2018general}, the authors explore a slightly different decomposition of the same MBO problem~\eqref{eq:MBO}. In particular, the authors let $f_1(\bar{x}) = \varphi(u) + \iota_{\bar{\mathcal{X}}\cup \{G \bar{x} = b\}}(\bar{x}) + q(\bar{x}) $, while $f_0(x) = 0$. In this way, the QUBO problem (first block update) becomes a projection problem of dimension $n$ onto the one dimensional constraint $\{0,1\}$, which is easily solvable, while the convex problem (second block update) becomes the convex relaxation of the MBO problem (with an additional penalization term). This non-convex ADMM heuristic is effective in finding approximate solutions to a wide variety of problems in classical computation, depending on an appropriate setting of the initial parameters. However, it is not readily applicable on quantum devices, as it does not involve QUBOs. 
\end{remark}

\subsection{From two-block to three-block ADMM for MBOs}

To overcome the limitation imposed by the convergence theorems (Theorem~\ref{th:wang:2}-\ref{th:wang:3}) on the smoothness of function $f_1(\bar{x})$, we use the same approach explored in Example~\ref{ex:2}, as well as in~\cite{Jiang2019}. We exploit a three-block implementation of ADMM (\threeadmm) onto the soft-constrained problem $(P'''):$
%
%
\begin{subequations}\label{eq:MBO:three-final}
\begin{eqnarray}
\minimize_{x \in \mathcal{X}, \bar{x} \in \mathbb{R}^m, y \in \mathbb{R}^n} \ &&  f_0(x) + f_1(\bar{x}) + \frac{\beta}{2}\|y\|^2_2  \\
\mathrm{subject~to:} && A_0 x + A_1\bar{x} = y,
\end{eqnarray}
\end{subequations}
where the only difference with~\eqref{eq:MBO:final} is the introduction of variable $y$, which penalizes constraint violations. 

Algorithm~\ref{algo:threeblock} reports the \threeadmm~algorithm, along with stopping criteria and evaluation metrics. As we can see, once again, the problem~\eqref{eq:MBO:three-final} is split into a QUBO, that can be solved by a QUBO oracle, and convex optimization problems.  We note that the two-block implementation is a particular case of the three-block algorithm, with $y_0 = 0 \in \mathbb{R}^n,$ and skipping third block update (i.e., step 4 of \ref{algo:threeblock}).

We are now ready for the convergence results for Algorithm~\ref{algo:threeblock} (\threeadmm). First, we present the results when continuous variables are not present, and then extend it to continuous variables. 

\begin{theorem}[Convergence of Algorithm~\ref{algo:threeblock}]\label{th.3}
Consider Problem~\eqref{eq:MBO} with no continuous variable $u$ and let Assumption~\ref{as:1} hold. Define the augmented Lagrangian,
\begin{align}
\mathcal{L}_{\varrho}(x,\bar{x},y,\lambda) & = f_0(x) + \iota_{\mathcal{X}}(x) +\frac{c}{2}\|G x - b\|^2_2 + \nonumber\\
& +  f_1(\bar{x}) + \frac{\beta}{2}\|y\|^2_2 + \lambda^{\transp}(A_0 x + A_1 \bar{x} - y ) + \nonumber\\
& + \frac{\varrho}{2}\|A_0 x + A_1 \bar{x} - y \|^2_2. 
\end{align}
Then, Algorithm~\ref{algo:threeblock} converges subsequently for any sufficiently large $\varrho > \max\{\beta,c\}$, that is, starting from any $x_0, \bar{x}_0, y_0, \lambda_0$, it generates a sequence that is bounded, has at least one limit point, and that each limit point $(x^*,\bar{x}^*, y^*, \lambda^*)$ is a stationary point of $\mathcal{L}_{\varrho}$, namely, $0 \in \partial\mathcal{L}_{\varrho}(x^*, \bar{x}^*, y^*, \lambda^*)$.

In addition, if $f_1(\bar{x})$ is a Kurdyka-\L ojasiewicz (K\L) function \cite{Lojasiewicz1993,Bolte2007,Attouch2013}, then $(x_k, \bar{x}_k, y_k, \lambda_k)$ converges globally to the unique limit point $(x^*, \bar{x}^*, y^*, \lambda^*)$.
\end{theorem}

\paragraph{Proof} \emph{We are going to leverage the results of Theorem~\ref{th:wang:3} to prove Theorem~\ref{th.3}. In particular, we are going to check that all the assumptions in Theorem~\ref{th:wang:3} are satisfied and determine a necessary condition on how large $\varrho$ must be for the algorithm to converge. }

\emph{{\bf A1)} (Coercivity). Coercivity holds since $x$ lies in a bounded set, $h(y) = \frac{\beta}{2}\|y\|^2_2$ is quadratic, therefore coercive, and the same holds for $\bar{x}$.  }

\emph{{\bf A2)} (Feasibility). $Im(A) \subseteq Im(B)$ holds by direct computation, since $A = [A_0, A_1] = [I_n, -I_n, 0_{l\times l}]$ and $B = -I_n$.}

\emph{{\bf A3)} (Lipschitz sub-minimization paths) $A_0 = -B = I_n$, so trivially $\bar{M} = 1$ for $x$ and $y$. Consider now $\bar{x}$, since no continuous variables are present $A_1 = -I_n$ and $\bar{M} = 1$  trivially. }

\emph{{\bf A4)} (Objective $f$-regularity). $f_0 + \iota_{\mathcal{X}}(x)$ is lower semi-continuous, and $f_1$ is restricted prox-regular since the sum of a convex function and the indicator function of a convex set.}

\emph{{\bf A5)} (objective $h$-regularity). $h(y) = \frac{\beta}{2}\|y\|^2_2$ is Lipschitz differentiable with constant $\beta$, so A5 holds. }

\emph{As for $\varrho$, from the conditions in~\cite[Lemma~9]{sun2019two}, then $\varrho> \max\{\beta,c\}$. }

\emph{And to finish the proof: $f_0(x) + \iota_{\mathcal{X}}(x) +\frac{c}{2}\|G x - b\|^2_2 + \frac{\beta}{2}\|y\|^2_2 + \lambda^{\transp}(A_0 x + A_1 \bar{x} - y ) + \frac{\varrho}{2}\|A_0 x + A_1 \bar{x} - y \|^2_2$ is  a K\L\, function, since it is semi-algebraic, and $\mathcal{L}_{\varrho}(x,\bar{x},y,\lambda)$ is K\L\,  if $f_1(\bar{x})$ is K\L.} \hfill $\Box$

\vskip2mm 

Theorem~\ref{th.3} describes a set of assumptions for which Algorithm~\ref{algo:threeblock} is proven to converge to a stationary point of the augmented Lagrangian $\mathcal{L}_{\varrho}$, which is a soft-constrained version of the original MBO problem~\eqref{eq:MBO}. We now expand on Theorem~\ref{th.3} by considering continuous variables.

		{\begin{algorithm}[H]
		\caption{\threeadmm~mixed-binary heuristic}\label{algo:threeblock} 
		\begin{algorithmic}[1]
			\small{\REQUIRE Initial choice of $x_0, \bar{x}_0, y_0, \lambda_0$. Choice of $\varrho, \beta, c >0$, tolerance $\epsilon>0$, and maximum number of iterations $K_{\text{max}}$.
				\WHILE {$k<K_{\text{max}}$ \AND $\|A_0 x^k+
					A_1 \bar{x}^k - y_k\| > \epsilon,$}
				\STATE First block update (QUBO):
				\begin{align}
				x_k = & \argmin_{x \in \{0,1\}^n} \,\, q (x) + \frac{c}{2}\|Gx - b\|^2_2 + \nonumber\\
				& + \lambda_{k-1}^{\transp} A_0 x + \frac{\varrho}{2}\|A_0 x + A_1 \bar{x}_{k-1} - y_{k-1} \|^2
			\end{align}
				\STATE Second block update (Convex):
				\begin{align}
				\bar{x}_k & = \argmin_{\bar{x} \in \mathbb{R}^m } \,\, f_1(\bar{x}) + \lambda_{k-1}^{\transp} A_1 \bar{x} + \nonumber\\
				&  \frac{\varrho}{2}\|A_0 x_{k} + A_1 \bar{x} - {y}_{k-1}\|^2
				\end{align}
				\STATE Third block update (Convex+quadratic):
				$$
				y_k = \argmin_{y \in \mathbb{R}^n } \,\, \frac{\beta}{2}\|y\|^2_2 - \lambda_{k-1}^{\transp} y + \frac{\varrho}{2}\|A_0 x_{k} + A_1 \bar{x}_{k} - {y}\|^2
				$$
				\STATE Dual variable update:
				$$
				\lambda_k = \lambda_{k-1} + \varrho(A_0 x_{k} + A_1 \bar{x}_k - y_k)
				$$
								\STATE Compute merit value:
\begin{align}
				\eta_k & = q(x_k)+\phi(\bar{x}_k) + \nonumber\\
				& + \mu (\max (g(x_k), 0) + \max(l(x_k, \bar{x}_k), 0))
\end{align}
				\ENDWHILE
			 \RETURN $x_{{k}^*}, \bar{x}_{{k}^*}, y_{{k}^*},$ with ${{k}^*} = \min_k \eta_k.$
		}
	\end{algorithmic}\end{algorithm}
	%

\begin{theorem}[Convergence of Algorithm~\ref{algo:threeblock} with continuous variables]\label{th.4}
The same results of Theorem~\ref{th.3} hold if:
\begin{itemize}
\item The function $\varphi(u)$ is strictly convex and the inequality constraint $\ell(z,u) \leq 0$ is never active, i.e., for each $z_k$ and $u_k$ generated by the algorithm we have $\ell(z_k,u_k) < 0$;
\item The inequality constraint $\ell(z,u) \leq 0$ is always active, i.e., for each $z_k$ and $u_k$ generated by the algorithm we have $\ell(z_k,u_k) = 0$, and for any fixed $z$, the inverse mapping $u(z) = \{u| \ell(z,u) = 0\}$ is unique and Lipschitz, i.e., $\|u(z) - u(z')\| \leq C \|z-z'\|$, for $C< \infty$, and $\varrho> \max\{C^2 \beta, c+C^2\}$.
\end{itemize}
\end{theorem}

\paragraph{Proof} \emph{We have only to show that A3 holds in these cases. For the first case, the inequality constraint is redundant and $u_k$ is only determined from $\varphi(u)$. Since $\varphi(u)$ is strictly convex, $u_k$ is unique and the same for all $k$'s, so $\|u_{k_1} - u_{k_2}\| = 0$ and A3 holds. This is the case, e.g., when inequality constraints are absent. }

\emph{For the second case, since $\ell(z_k,u_k) = 0$ and $\|u(z) - u(z')\| \leq C \|z-z'\|$, then A3 holds with $\bar{M} = C$. And the conditions on $\varrho$ are derived from~\cite[Lemma~9]{sun2019two}. This is the case, e.g., when the inequalities are linear equality constraints as $F z + H u = g$, and $H$ is full rank.}\hfill $\Box$

The conditions of Theorems~\ref{th.3} and~\ref{th.4} are quite mild in many practical relevant MBO problems. In full generality however, Algorithm~\ref{algo:threeblock} is a heuristic algorithm, especially as we remark next.

\begin{itemize}
\item {\bf Equality constraints}. When equality constraints are presents, they are softened with the augmented term $\frac{c}{2}\|G x - b\|^2_2$ in the cost function. This induces a trade-off: from the conditions in~\cite[Lemma~9]{sun2019two}, then at least $\varrho> \max\{\beta, c\}$; however, to enforce the equality constraints, these have to be at least as important as the enforcing of zero residuals, i.e., $c\geq \varrho$. This introduces the trade-off of either terminating with a solution with zero residuals (meaning the convergence has been reached, but equality constraints are not necessarily satisfied), or with equality constraints satisfied (without bounds on the magnitude of the residual). 

Note that off-loading the equality constraints to variable $\bar{x}$ and imposing them exactly, only mildly solves the issues, since residual convergence would be achieved with $y \neq 0$ (in general) and therefore the equality constraints will not be satisfied exactly. 

\item {\bf Continuous variables}. When continuous variables are present, which do not satisfy either of the conditions of Theorem~\ref{th.4}, then assumption A3 is not satisfied, making Theorem~\ref{th.3} not hold in this situation and Algorithm~\ref{algo:threeblock} is still a heuristic for this case. 

\end{itemize}



\subsection{Inexact optimization and noise}\label{sec:inexact}

\rev{
We briefly remark here the effect of inexact optimization of the binary subproblems as well as noise in real quantum devices. The theoretical analysis assumes that the QUBOs need to be solved exactly to guarantee the validity of the presented theorems. In practical situations however, this is hard to achieve and this requirement needs to be weakened. For instance, in current noisy quantum computers, noise is inherent in the computations and an exact optimization is rather far-fetched. Even in the case of noise-free quantum computers, solving QUBOs at optimality may be unrealistic, especially in large-scale instances. Withing the ADMM framework, it then makes sense to ask \emph{(i)} whether one can tolerate inexact optimization; \emph{(ii)} whether noise has to be always detrimental for convergence. }

\rev{
On the first point, we do not have an answer yet in the general case (even though this has been studied abundantly and with positive answers in the convex case). In~\cite{wang2019global}, the authors show that one can tolerate inexact computations which are asymptotically vanishing and summable. In particular, if one can solve the QUBOs with increasing degree of accuracy while the algorithm progresses, then convergence can be still established. This is a promising first result.}

\rev{The second point is even more interesting and open. It is fairly understood and it has been experimentally observed that a small amount of noise can help to ``guide'' convergence of first-order algorithms to global optimizers in non-convex problems (e.g., by escaping local minima, or saddle-points). So, it is possible that a small amount of noise, inexactness, or both, could help convergence instead of jeopardizing it. We will explore this aspect in the simulation results}.

\section{Simple examples}\label{sec:simple-ex}

We discuss here some interesting examples to showcase the performance of \twoadmm~and \threeadmm~for MBOs problems in simple settings, and gain some insights on the solutions obtained.

\subsection{Inequaltity constraints}

\begin{example}\label{ex:3}
Consider the problem:
\begin{eqnarray}
\min_{x \in \{0,1\}^2} && v + w,\\ \textrm{subject to: } && 2 v + w \leq 2,\\ && v + w \geq 1,  
\end{eqnarray}
where $x = [v, w]^{\transp}$. We consider two cases, Case 1: $1001 = \varrho > \beta = 1000$ (verifying the necessary conditions for Algorithm~\ref{algo:threeblock} to converge, but Algorithm~\ref{algo:twoblock} is a heuristic), and Case 2: $\varrho = \beta = 1000$, for which both algorithms are heuristics. Figure~\ref{fig:ex-3} showcases convergence of the residual of both Algorithm~\ref{algo:twoblock} and Algorithm~\ref{algo:threeblock}, where we defined the three-block residual as $r_3 = \|A_0 x + A_1 \bar{x} - y \|$, while the restricted three-block residual as $rr_3 = \|A_0 x + A_1 \bar{x}\|$ (that is how far we are from the solution of the non-relaxed problem), as well as the two-block residual as $r_2 = \|A_0 x + A_1 \bar{x}\|$. 

As we can see, in Case 1, Algorithm~\ref{algo:threeblock} converges in the residual sense while Algorithm~\ref{algo:twoblock} does not. In particular, the results (at three significative digits) yield: $x = [0, 0] , \bar{x} = [0.499, 0.500], y = [-0.499, -0.499]$ for the three-blocks, while $x = [0, 1], \bar{x} = [0.499, 0.999]$, for the two blocks. We can also see that, despite Algorithm~\ref{algo:threeblock} convergence, the result $x = [0, 0]$ is not optimal (not even feasible for the original non-relaxed problem), while Algorithm~\ref{algo:twoblock} delivers one of the two optimal results $x = [0, 1]$, while not converging. 

In Case 2, Algorithm~\ref{algo:threeblock} converges in the residual sense (even though it is not guaranteed to do so) while Algorithm~\ref{algo:twoblock} does not. In particular, the results (at three significative digits) yield: $x = [0, 1] , \bar{x} = [0.002, 0.999], y = [-0.002, -0.000]$ for the three-blocks, while $x = [0, 1], \bar{x} = [0.499, 0.999]$, for the two blocks. In this case, both Algorithm~\ref{algo:threeblock} and Algorithm~\ref{algo:twoblock} deliver one of the two optimal results $x = [0, 1]$.  
\end{example}

\begin{figure}
\centering
\includegraphics[width = 0.48\textwidth]{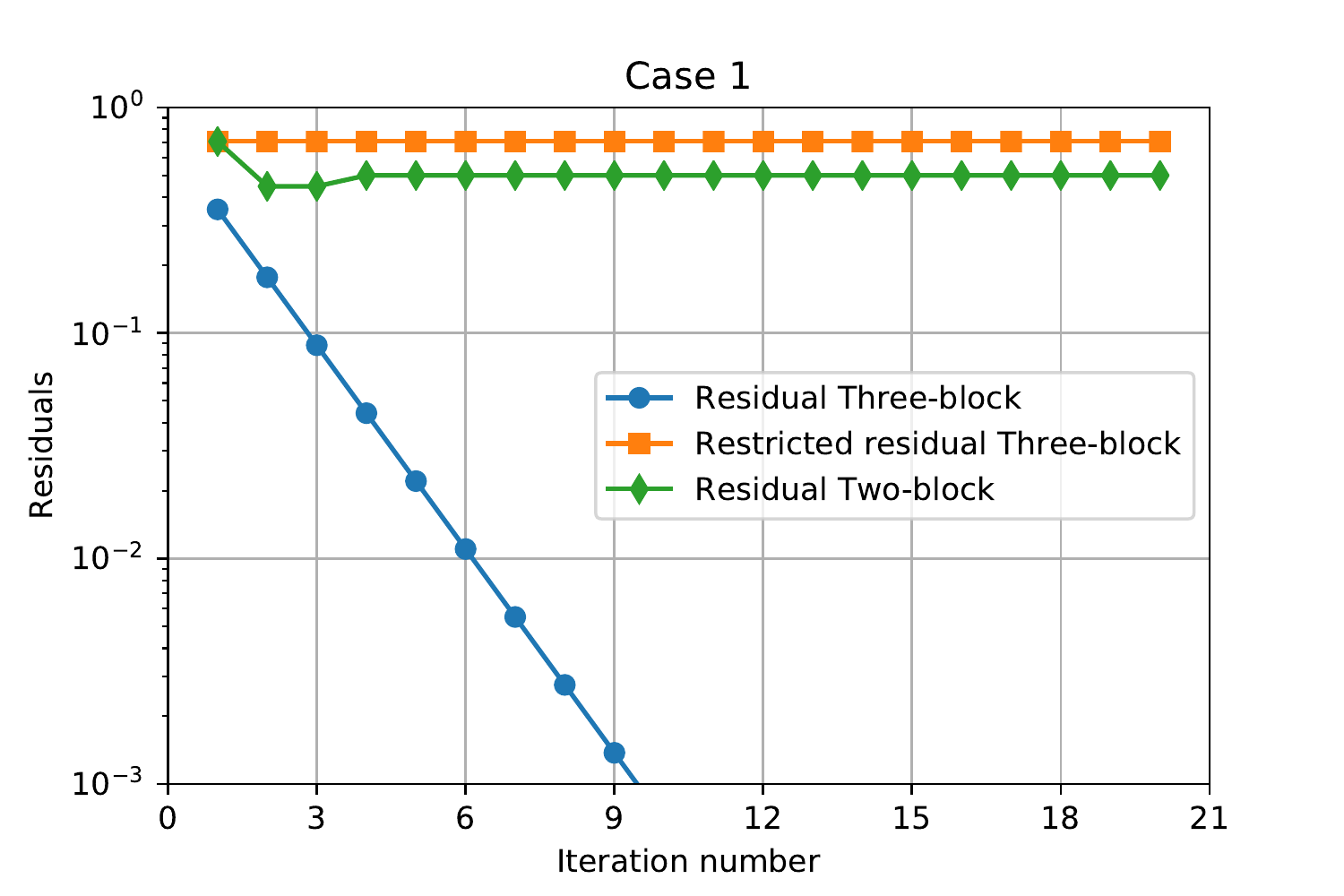}
\includegraphics[width = 0.48\textwidth]{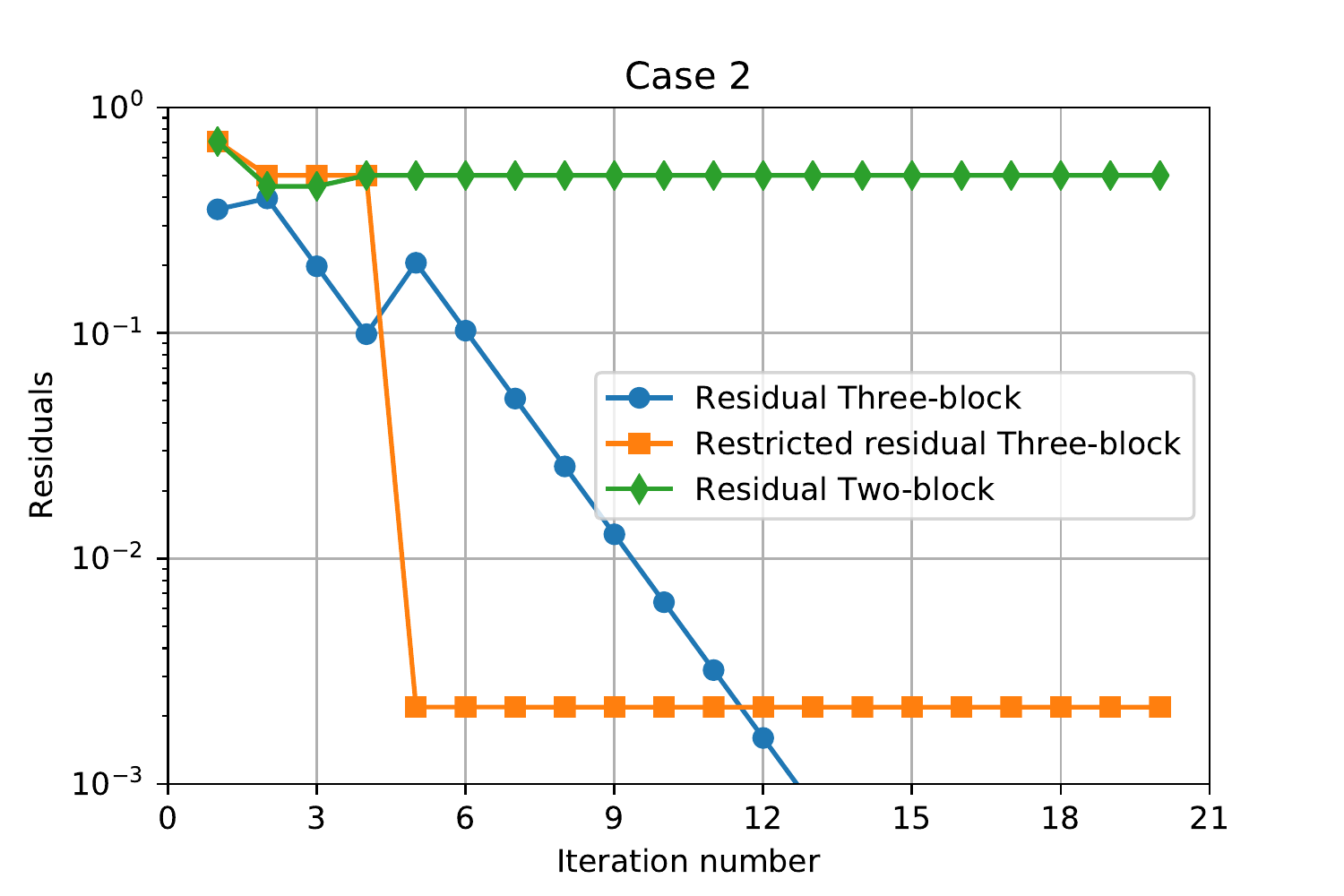}
\caption{Convergence of the residuals for Example~\ref{ex:3}.}
\label{fig:ex-3}
\end{figure}

\begin{example}\label{ex:4}
Consider the problem:
\begin{eqnarray}
\min_{x \in \{0,1\}^3} &&  v + w + t, \\ \textrm{subject to: } && 2 v + 10 w + t \leq 3, \\ && v + w + t \geq b
\end{eqnarray}
where $x = [v, w, t]^{\transp}$ and $b$ is either $1$ (Case 1) or $2$ (Case 2). We fix $\varrho = 1001, \beta = 1000$. 

In Case 1, as we can see in Figure~\ref{fig:ex-4}, both algorithms converge. Algorithm~\ref{algo:threeblock} delivering $x = [0., 0., 0.]$, $\bar{x} =  [0.397, 0.178, 0.424]$, $y = [-0.397, -0.178, -0.424]$; Algorithm~\ref{algo:twoblock} delivering a feasible solution $x = [1., 0., 1.]$, $\bar{x} = [ 1., 0., 1.]$. 

In Case 2, both algorithms converge and deliver the optimal solution. 
\end{example}

\begin{figure}
\centering
\includegraphics[width = 0.48\textwidth]{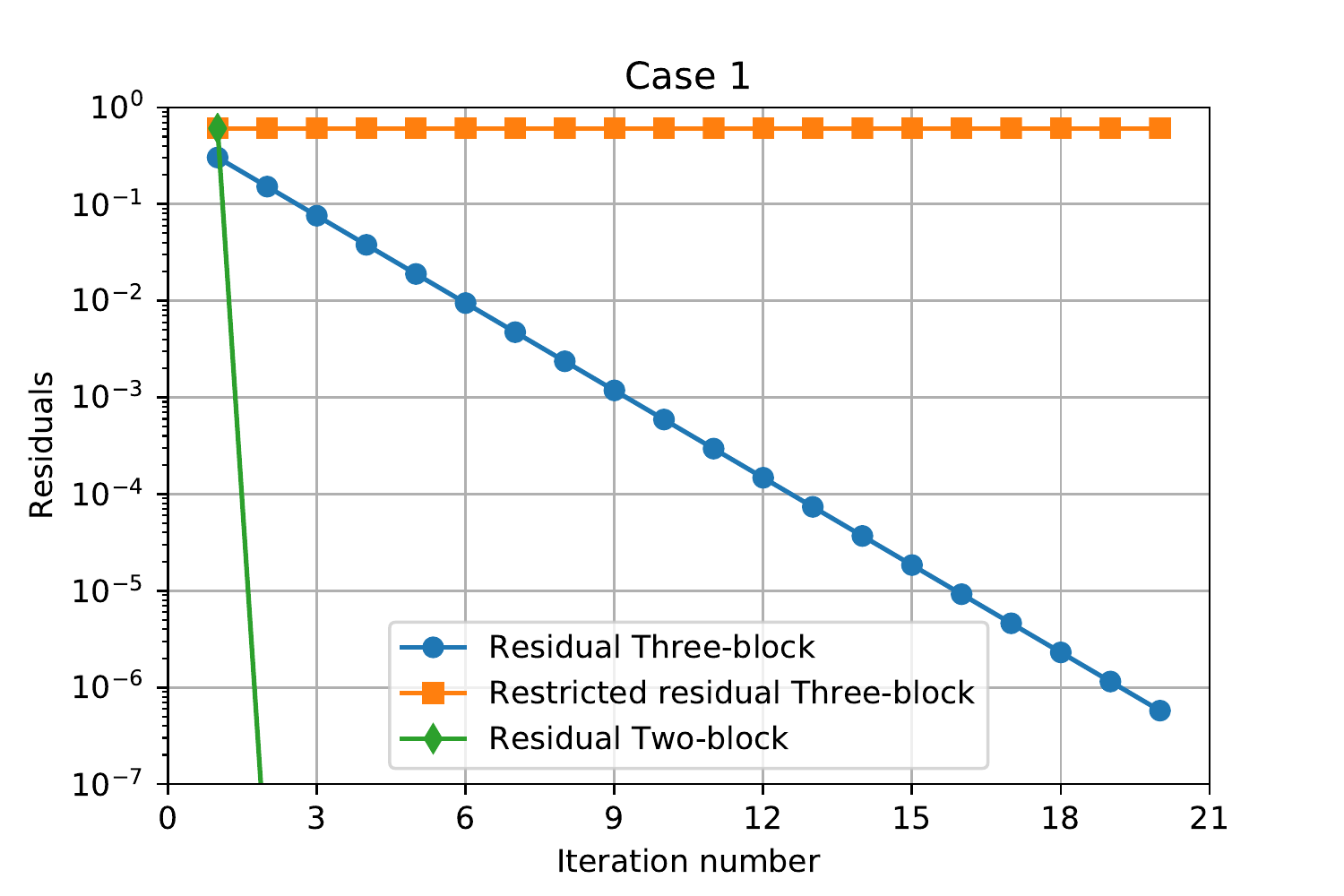}
\includegraphics[width = 0.48\textwidth]{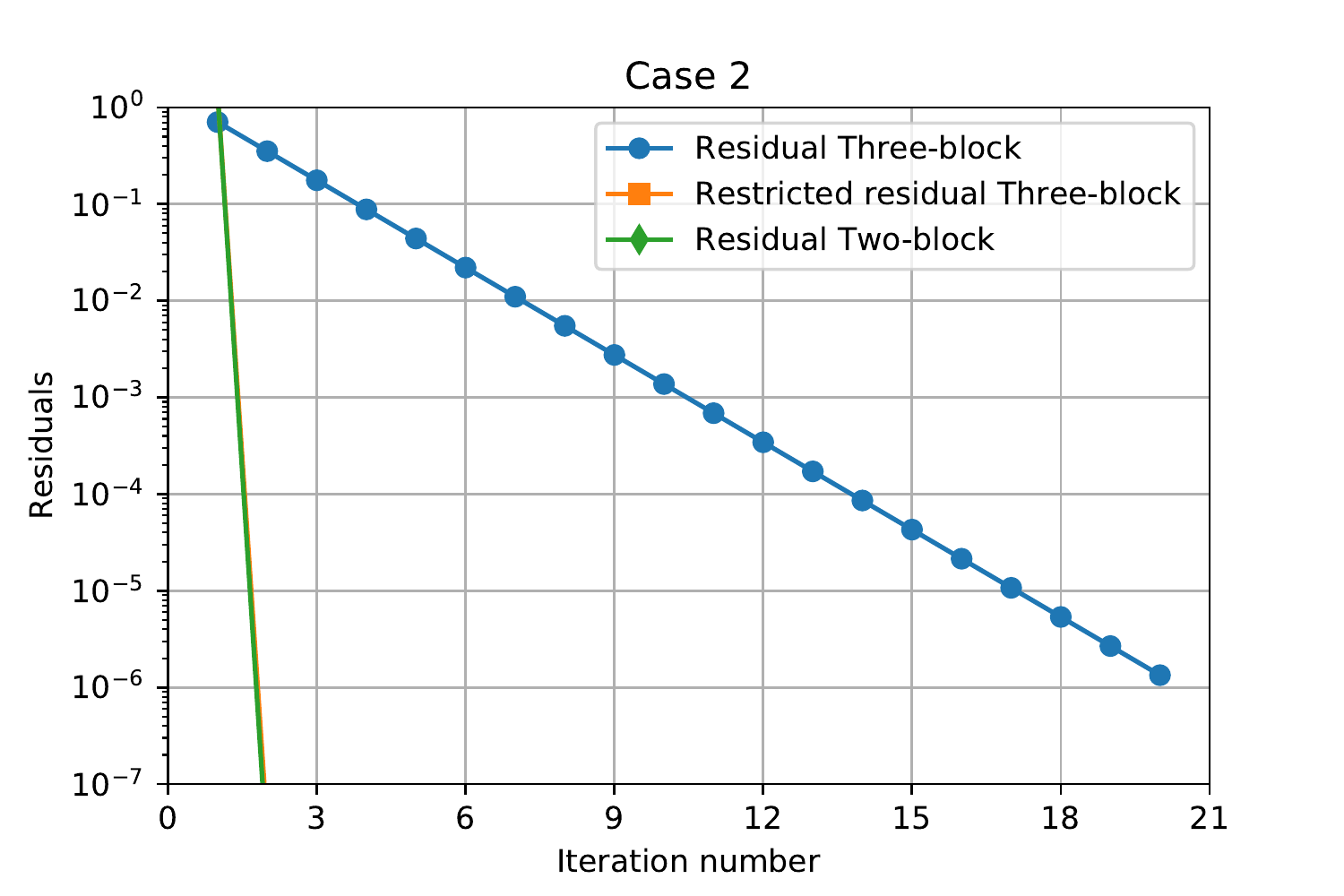}
\caption{Convergence of the residuals for Example~\ref{ex:4}.}
\label{fig:ex-4}
\end{figure}

\subsection{Equalities and Inequaltity constraints}

\begin{example}\label{ex:5}
Consider the problem:
\begin{eqnarray}
\min_{x \in \{0,1\}^3} && \,  v + w + t,\\ \textrm{subject to: }&&  2 v + 2 w + t \leq 3, \\ && v + w + t \geq 1, \\&& v + w = 1,
\end{eqnarray}
where $x = [v, w, t]^{\transp}$. We fix $\varrho = 1001, \beta = 1000$, and the penalization parameter for the equality constraint to be $c = 900$ (Case 1), $c=1100$ (Case 2). 

In Case 1, Algorithm~\ref{algo:threeblock} is supposed to converge. Both algorithms converge in practice, Algorithm~\ref{algo:threeblock} to one optimal solution $x = [1., 0., 0.]$, $\bar{x}= [1., 0.002, 0.002]$, $y=[0., -0.002, -0.002]$; Algorithm~\ref{algo:twoblock} to a feasible solution $ x= [1., 0., 1.]$, $\bar{x}= [0.999, 0., 0.999]$. 

In Case 2, Algorithm~\ref{algo:threeblock} is not guaranteed to converge. However, both algorithms seem to converge. Both deliver an optimal solution:  $x = [0., 1., 0.]$.
%
%
\end{example}

\begin{figure}
\centering
\includegraphics[width = 0.48\textwidth]{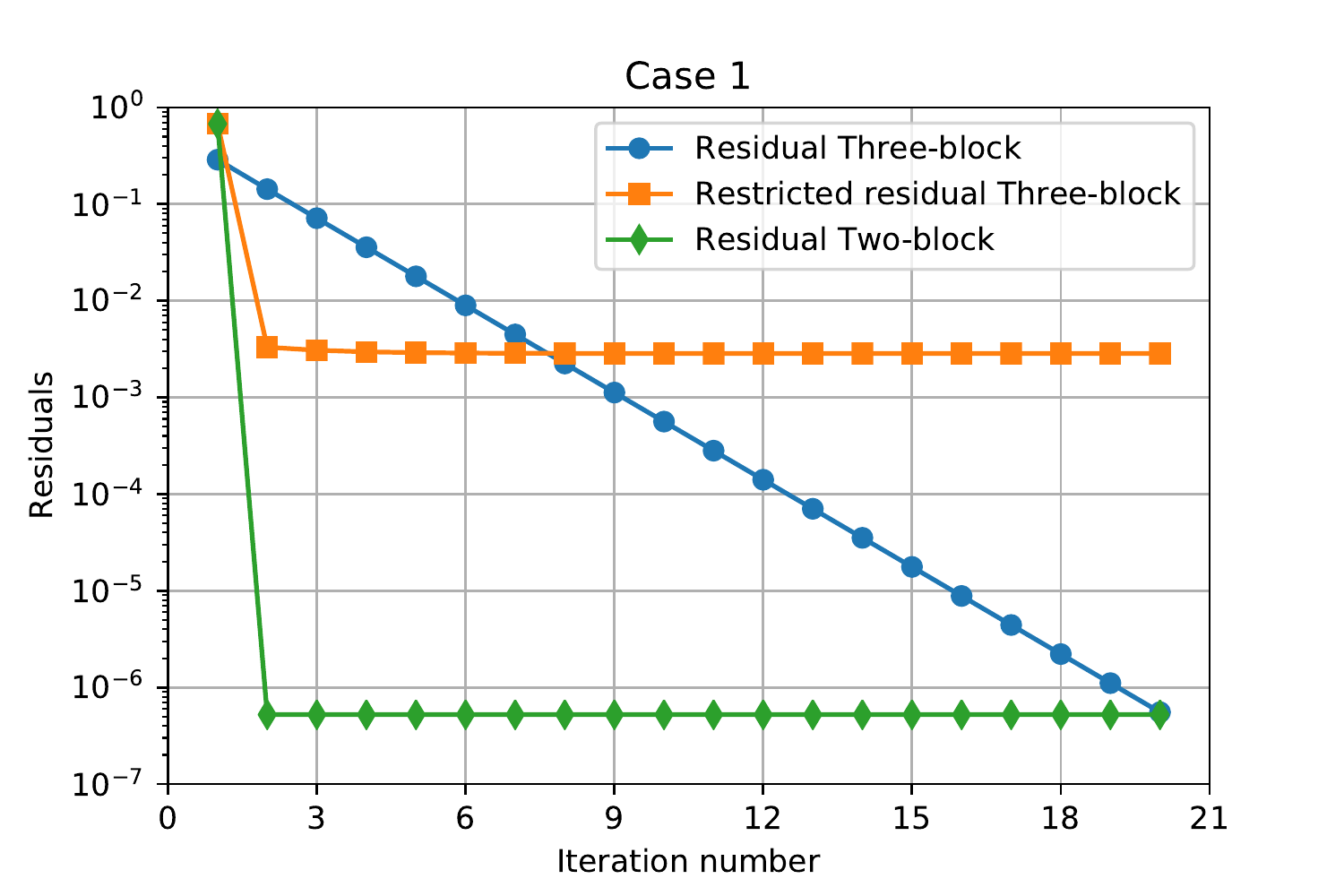}
\includegraphics[width = 0.48\textwidth]{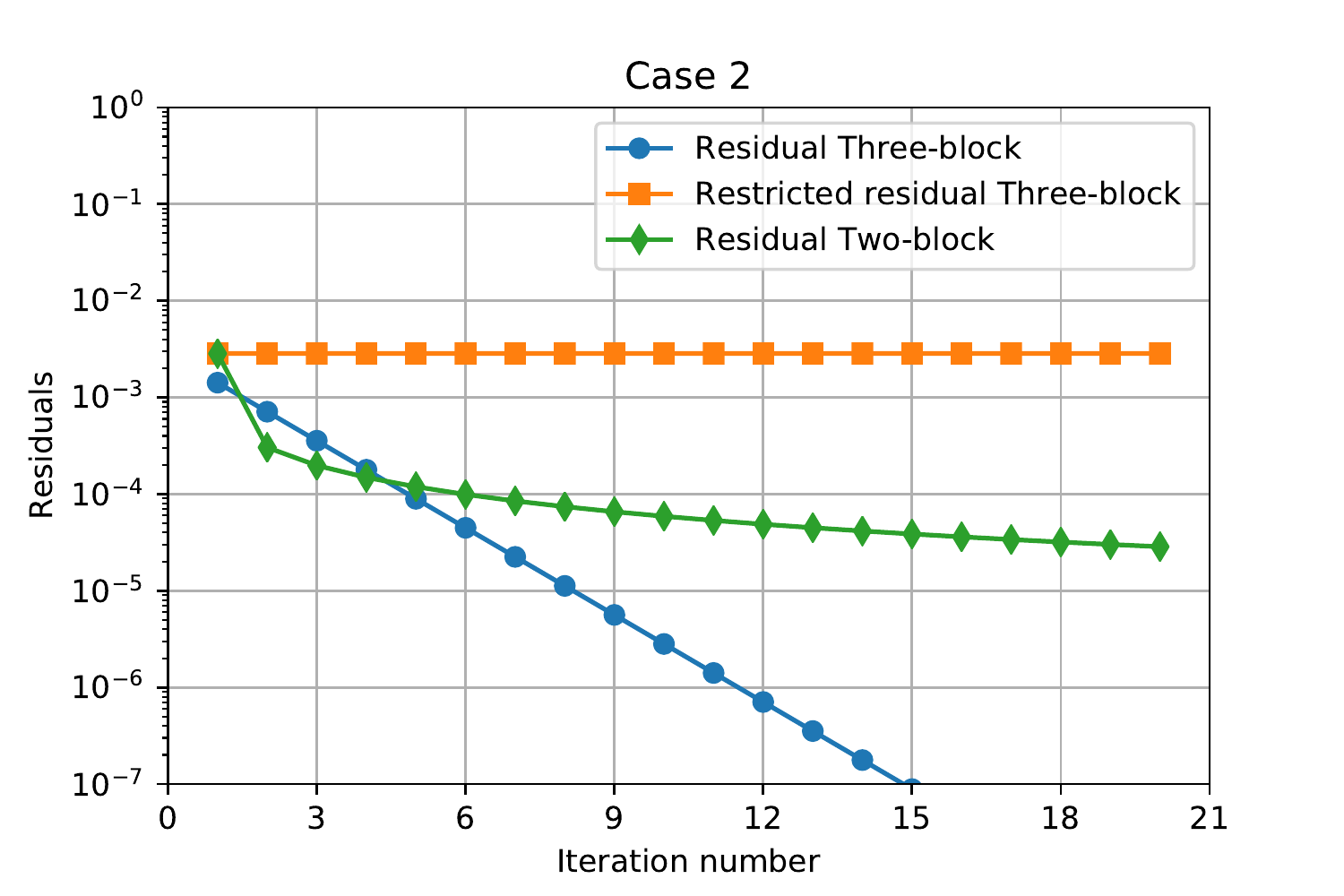}
\caption{Convergence of the residuals for Example~\ref{ex:5}.}
\label{fig:ex-5}
\end{figure}

\subsection{Continuous variables}

\begin{example}\label{ex:6}
Consider the problem:
\begin{eqnarray}
\min_{x \in \{0,1\}^3, u\in\mathbb{R}} && \,  v + w + t + 5 (u - 2)^2,\\ 
\textrm{subject to: }&&  v + 2 w + t + u \leq 3, \\
&&  v + w + t \geq 1, \\ && v + w = 1,
\end{eqnarray}
where $x = [v, w, t]^{\transp}$. We fix $\varrho = 1001, \beta = 1000$, and the penalization parameter for the equality constraint to be $c = 900$. The inequality constraint with the continuous variable is always active, so Algorithm~\ref{algo:threeblock} is supposed to converge (as for Theorem~\ref{th.4}). 

In Figure~\ref{fig:ex-6}, we see how both algorithms converge, but only Algorithms~\ref{algo:threeblock} yield the optimal solution (incurring zero optimality gap). In particular, Algorithm~\ref{algo:threeblock} delivers the optimal solution $x = [1.,0.,0.]$, $\bar{x} = [1., 0., 0., 2.]$, $y = [0., 0., 0.]$, whereas Algorithm~\ref{algo:twoblock} delivers the feasible solution $x = [1.,0.,1.]$, $\bar{x} = [1., 0., 1., 1.]$.
\end{example}

\begin{figure}
\centering
\includegraphics[width = 0.48\textwidth]{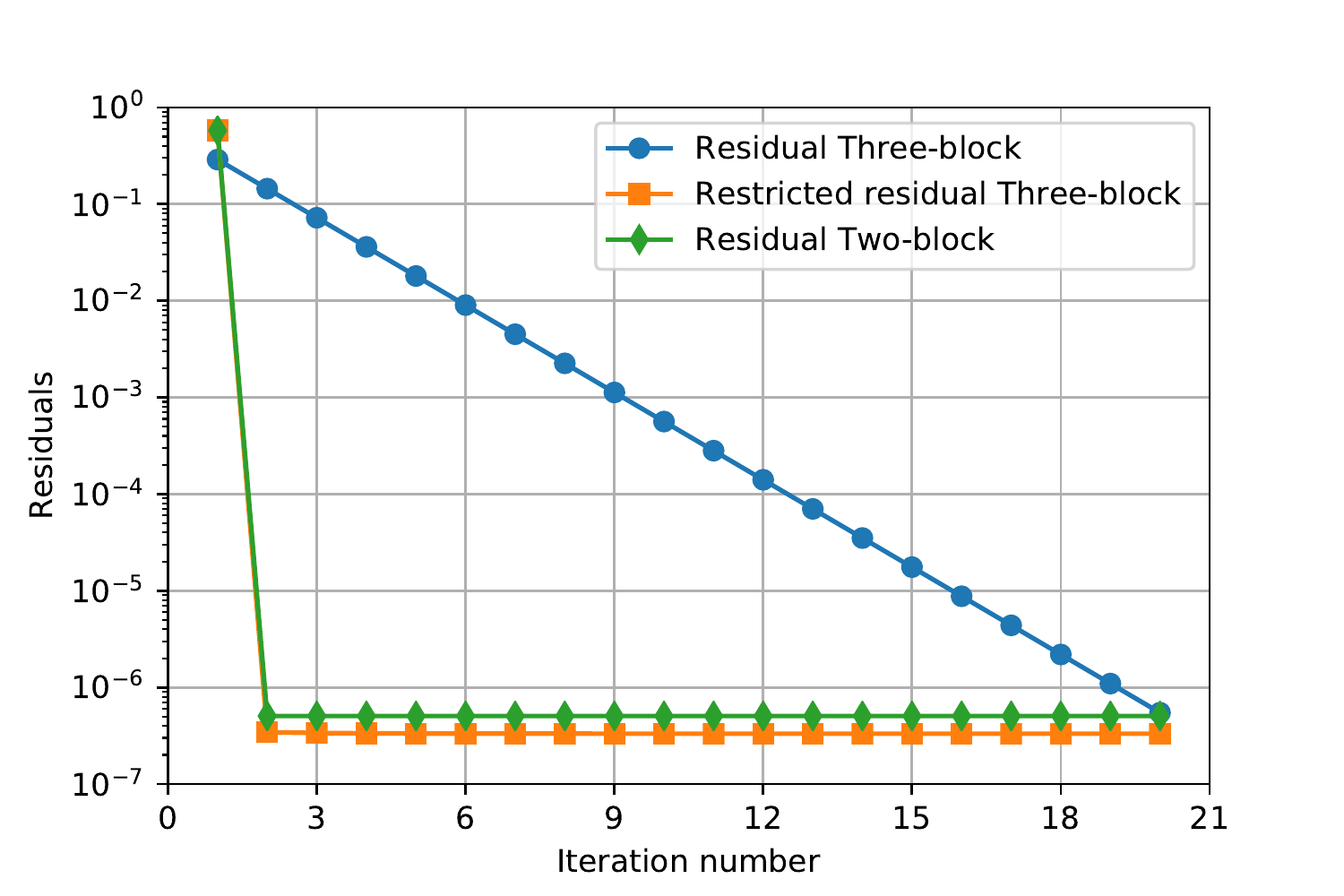}
\includegraphics[width = 0.48\textwidth]{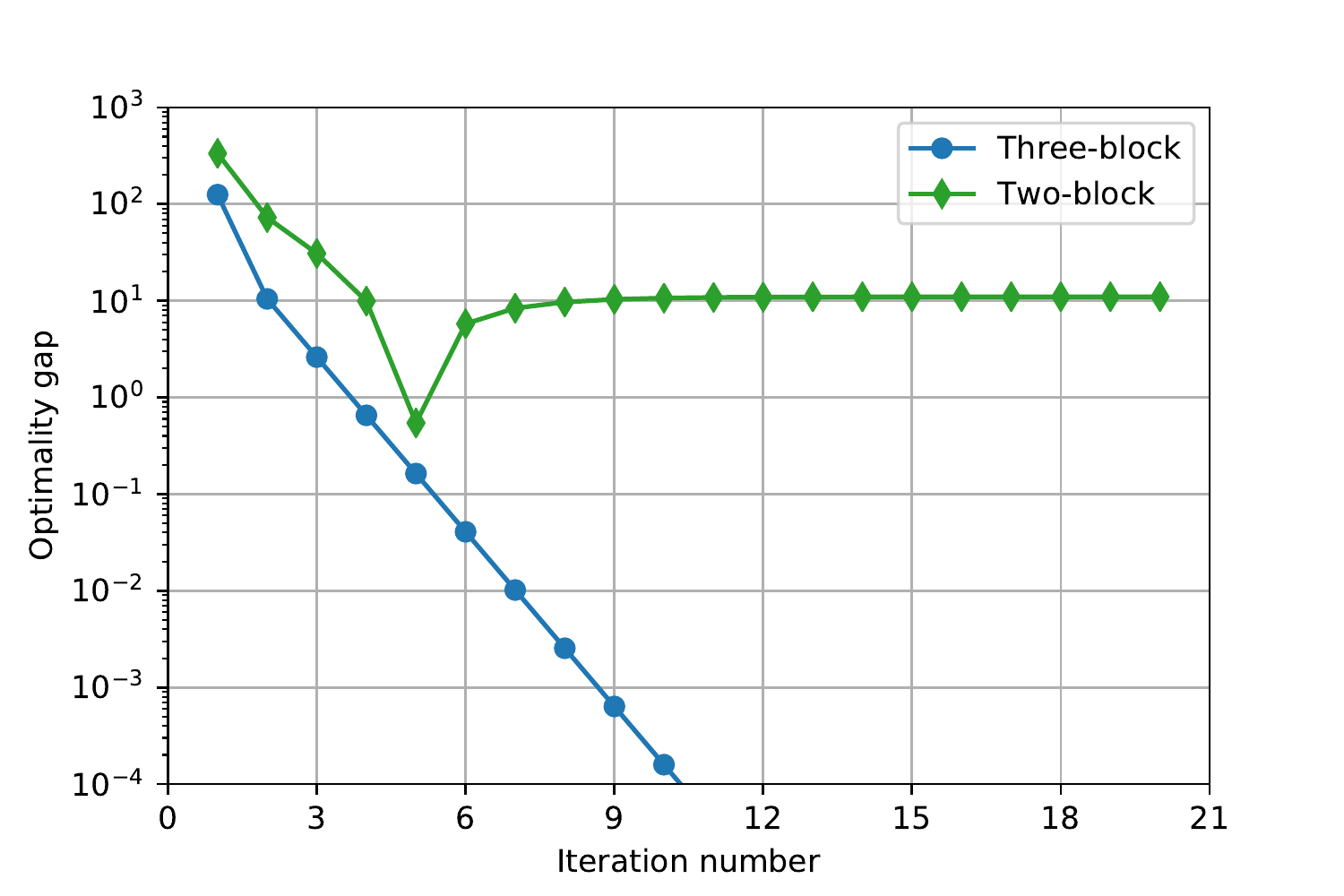}
\caption{Convergence of the residuals and optimality gap for Example~\ref{ex:6}.}
\label{fig:ex-6}
\end{figure}

\subsection{Inexact updates}

\begin{example}\label{ex:7}
\rev{We re-consider now Example~4, in Case 1. There, as we could see in Figure~\ref{fig:ex-4}, both algorithms converged, but Algorithm~\ref{algo:threeblock} was delivering an infeasible solution. We consider here the case in which the QUBO is solved with some errors, and we model these errors as probability of a bit-flip of the QUBO optimal solution. In particular, for each component $x_{k(i)}$ of the QUBO solution, with value either $0$ or $1$, we consider that there is a certain probability for it to flip to $1$ or $0$, respectively. We also consider that this probability decreases as $50/k$\% as the number of iterations $k$ increases, to model the fact that we are solving the QUBO subproblems more and more accurately. This allows \admm~to escape bad regions of the solution space at first, and to intensify the search for higher-quality solutions in the consequent iterations.}

\rev{In Figure~\ref{fig:ex-7}, we see how both algorithms converge, but now Algorithm~\ref{algo:threeblock} yields an optimal solution $x = [1., 0., 0.]$, $\bar{x} =  [1., 0.002, 0.002]$, $y = [0.0, -0.002, -0.002]$; Algorithm~\ref{algo:twoblock} delivers the same feasible solution as before. }

\rev{This example showcases how noise in real setting can help the algorithms to converge to optimal solutions. Note that in Case 2, the same optimal solution is achieved also with noise in this setting. 
}
\end{example}

\begin{figure}
\centering
\includegraphics[width = 0.48\textwidth]{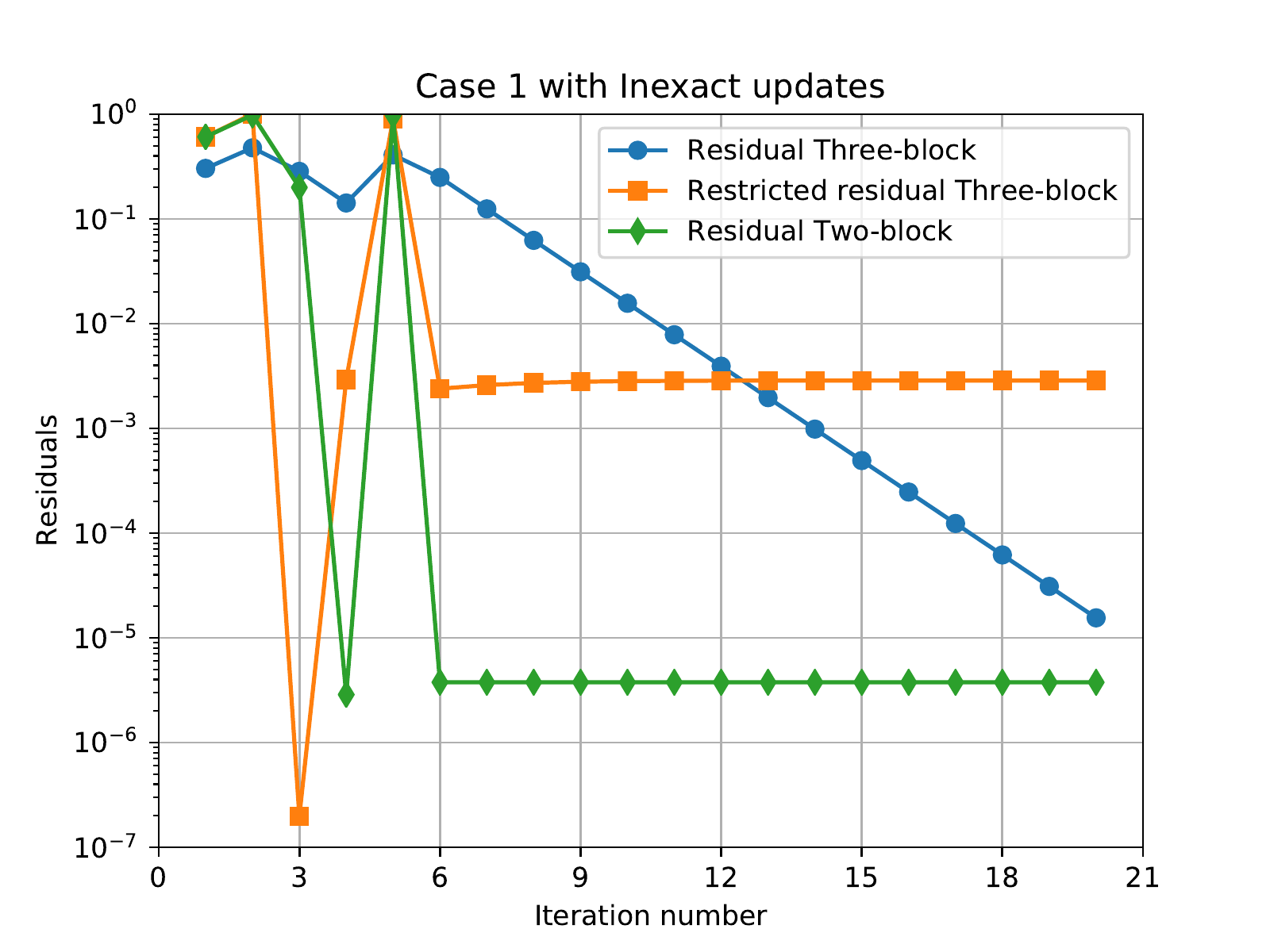}
\caption{Convergence of the residuals for Example~\ref{ex:7}.}
\label{fig:ex-7}
\end{figure}

\section{Mixed-Binary Optimization problems}\label{sec:MBO}

As shown in the simple examples presented in \Cref{sec:simple-ex}, \threeadmm~(Algorithm~\ref{algo:threeblock}) and \twoadmm~(Algorithm~\ref{algo:twoblock}) are heuristics, in the general case. In the best case, \threeadmm~is guaranteed to converge and it delivers an optimal solution for the original MBO. In the worst case, both algorithms fail to deliver feasible solutions. In the middle, \threeadmm~may converge, but the soft-constrained solution is not feasible with respect to the hard-constrained formulation, or both algorithms could converge to a feasible but not optimal solution. 
With this in mind, we are now ready to apply the algorithms to two well-known MBO problems: Bin Packing (BP) Problem and Mixed Integer Setup Knapsack (MISK) problem. The computational results will be discussed in \Cref{sec:comp-res}, where we will show that despite the heuristic nature of \admm, we can still obtain feasible solutions in many cases. This is not trivial in general for combinatorial optimization problems \cite{fischetti2005feasibility, bertacco2007feasibility}.

The BP is arguably one of the most studied combinatorial problems \cite{DELORME20161}. Being strongly NP-hard, it stimulated the study of heuristics, metaheuristics and worst-case approximation bounds. 
Given $n$ items, each having an integer weight $w_j, j = 1, \dots, n$, and $m$ identical bins of integer capacity $Q$, the aim of BP is to pack all the items into the minimum number of bins so that the total weight packed in any bin does not exceed the capacity. Applications of BP in logistics and scheduling are numerous, and include cutting stock problems, containers loading, data storage, job scheduling and  resource allocation. \\
The MISK belongs to the class of Knapsack Problems \cite{martello1990knapsack, kellerer2013knapsack}. The classical knapsack problem is that of deciding which items to pack in a capacitated knapsack, so as to maximize the profit of the items in the knapsack. In the setup knapsack problem (SKP), each item belongs to a family, and an item can be assigned to the knapsack only if a setup charge for the correspondent family is paid \cite{lin1998bibliographical}. SKP can model capacitated scheduling problems. In the MISK, items can be fractionally assigned to the knapsack. MISK appears as a subproblem of the capacitated coordinated replenishment problem.

\subsection{Binary Linear Programming Formulation for Bin Packing}
Let $\xi_{ij} \in \{0,1\}$ be the binary decision variable which, if $1$, indicates that item $j$ is assigned to bin $i$. Let $\chi_i \in \{0,1\}$ be the binary decision variable which, if $1$, indicates that bin $i$ is containing items.
A natural mathematical formulation for Bin Packing (BP) problem is then given by the binary linear program:
\begin{subequations}
\begin{align}
\minimize_{\chi, \xi} \quad& \displaystyle\sum_{i=1}^m \chi_i   \label{BPP:obj}\\
\text{subject to:} \quad & \displaystyle\sum_{i=1}^m \xi_{ij} =1 \qquad j =1, \dots, n \label{BPP:assign}\\
	& 	\displaystyle\sum_{j=1}^n w_j \xi_{ij} \leq Q \chi_i \qquad i = 1, \dots, m \label{BPP:capacity}\\
	& \xi_{ij} \in \{0,1\} \qquad i = 1, \dots, m,\nonumber \\
	 & \phantom{\xi_{ij} \in \{0,1\}} \qquad j =1, \dots, n \label{BPP:x}\\
	& \chi_{i} \in \{0,1\} \qquad i = 1, \dots, m \label{BPP:y}
\end{align}
\end{subequations}
In particular:
\begin{itemize}
	\item The objective function \eqref{BPP:obj} is the number of bins in solution.
	\item Constraints \eqref{BPP:assign} enforce the assignment of each item into a bin.
	\item Constraints \eqref{BPP:capacity} ensure the packed items do not exceed the bin capacity.
	\item Constraints \eqref{BPP:x} and \eqref{BPP:y} express the bounds on the decision variables.
\end{itemize}
The presence of inequalities to express the capacity constraints \eqref{BPP:capacity} forbids the straightforward mapping to an Ising Hamiltonian model, and the direct application of quantum optimization algorithms, such as VQE \cite{mcclean2016theory} and QAOA \cite{farhi2014quantum}.\\

\subsection{Mixed-Binary Formulation for the Mixed Integer Setup Knapsack (MISK)}
The MISK has received limited attention in literature. The mixed-integer formulation proposed in \cite{altay2008exact} is presented in this section. The items belongs to $K$ non-overlapping families. Each family $k$ has $T$ items, and a setup cost $S_k \geq 0$, when included in the knapsack. Each item $t$ of family $k$ has a value $C_{kt} < 0$, and a resource consumption $D_{kt} \leq 0$, if assigned to a knapsack with capacity $P$. The decision variables are the fraction $\xi_{kt}$ of item $t$ that is included in the knapsack, and is the binary decision $\chi_k$ to setup family $k$ in the knapsack. MISK can then be formulated as:
\begin{subequations}
\begin{align}
\minimize_{\chi, \xi}\quad & \displaystyle\sum_{k=1}^K S_k \chi_k + \sum_{k=1}^K \sum_{t=1}^T C_{kt} \xi_{kt}  \label{MISKP:obj}\\
\text{subject to:} \quad & \displaystyle\sum_{k=1}^K \sum_{t=1}^T D_{kt} \xi_{kt} \leq P  \label{MISKP:capacity}\\
& \xi_{kt} \leq \chi_k \qquad k = 1, \dots, K, t = 1, \dots, T \label{MISKP:allocation}\\
& \xi_{kt} \geq 0 \qquad k = 1, \dots, K, t = 1, \dots, T \label{MISKP:x}\\
& \chi_{k} \in \{0,1\} \qquad k = 1, \dots, K \label{MISKP:y}
\end{align}
\end{subequations}

The aim is to minimize the setup costs and maximize the value of the assigned items via the objective function \eqref{MISKP:obj}. Constraints \eqref{MISKP:capacity} ensures that the capacity of the knapsack is not violated: this is a fixed charge capacity constraints, because setup capacity consumption is not considered. Constraints \eqref{MISKP:allocation} impose that if item $t$ of family $k$ is assigned to the knapsack, then the setup cost of family $k$ is paid accordingly.

\section{Computational Results}\label{sec:comp-res}

We discuss here the multi-block (\admm)~results on BP and MISK. The algorithm has been implemented in Python on a  machine with 2.2 GHz, Intel Core i7 processor, and a RAM of 16 GB; the simulations on quantum devices to solve the QUBOs have been conducted by using the Qiskit framework \cite{qiskit-link} 
\rev{(specifically, \textit{qiskit} version 0.15.0, \textit{qiskit-aqua} version 0.6.1, 
\textit{qiskit-terra} version 0.10.0, 
\textit{qiskit-aer}  version 0.3.2)},
while IBM ILOG CPLEX 12.8 has been chosen as classical optimization solver 
\footnote{IBM, IBM Q, Qiskit are trademarks of International
	Business Machines Corporation, registered in many jurisdictions worldwide. Other product or service names
	may be trademarks or service marks of IBM or other
	companies.}.

In Figure~\ref{fig.appr}, a summary of the proposed approach and implementation choices are presented \revision{with VQE and QAOA as quantum QUBO solvers. It is important to note the presence of two nested iterations: the outer one due to ADMM and the inner one due to the classical solver for VQE/QAOA. In the following, we consistently call the ADMM iterations ``outer'' iterations, while the classical solver ones are ``inner'' iterations. The choice of SPSA or COBYLA affects the choice of number of inner iterations, therefore, we often say: SPSA/COBYLA inner iterations. Furthermore}, for all simulations reported in the following subsections:

\begin{itemize}
\item \textbf{The \admm\ algorithm} has been run with a time limit of $1$ hour, \revision{a limit of $500$ outer iterations}, and with merit parameter $\mu = 1e+3$.

In addition, to avoid large penalization factors $\varrho$ from the first \revision{outer} iteration, we start with the value $\varrho = 1e+4$, which is then increased by $10\%$ at each iteration, until it exceeds the value of $1e+7$. 
The penalization $c$ of equality constraints has been set to $1e+5$. The penalization $\beta$ of residual $\lVert y \rVert$ is initially set to $1e+3$ and then updated according to scheme described in \cite{sun2019two}, specifically $\beta^{k+1} = \gamma \beta^k$, if $\lVert \bar{x}^k \rVert \leq \omega \lVert \bar{x}^{k-1} \rVert$, with $\omega = 0.5$ and $\gamma = 2$, so to foster exact penalization. 

\item \textbf{The QUBO subproblems} are solved either classically with CPLEX, or on the simulated quantum devices via the Qiskit APIs. A common random seed has been fixed for all simulations. No limitations on the running time of the quantum solver have been imposed \revision{but only a maximum number of inner iterations}. 
The Variational Quantum Eigensolver (VQE) has been invoked with the RY variational form in a circuit of depth $5$ and full entanglement, and the QASM simulator as Qiskit Aer backend. Figure~\ref{fig.circuit} represents the circuit that was used in the case of three qubits and depth $4$. \rev{The Quantum Approximate Optimization Algorithm (QAOA) has been tested with circuit depth $3$ and the same backend.}

\item \textbf{VQE} is itself an iterative quantum algorithm that involves defining a parametrized variational form and optimizing classically on the rotation parameter vector $\theta$, while evaluating the variational form and its gradients on the quantum device. In our simulations, the classical solvers used by VQE are the model-based local optimizers Simultaneous Perturbation Stochastic Approximation (SPSA) \cite{spall1992multivariate}, and Constrained Optimization By Linear Approximation (COBYLA) \cite{gomez2013advances}. For both solvers, the Qiskit implementation has been used.
\item \rev{\textbf{QAOA} generalizes VQE because
	  the variational form is added with parameter vector $\beta$ of length equal to $\theta$. As for VQE, the classical optimization is performed via SPSA and COBYLA.}
  \item \revision{For the sake of clarity, we indicate the quantum QUBO solvers with name \textit{quantum algorithm}-\textit{internal classical solver}. For instance, VQE-SPSA solves QUBOs with VQE and SPSA as internal classical optimizer.  For the classical optimizers SPSA and COBYLA on the rotation parameters, sensitivity results are reported for $10, 20, 50$ maximum \revision{inner} iterations. }
\end{itemize}

\begin{figure*}[t]
\centering
\includegraphics[width=0.95\textwidth]{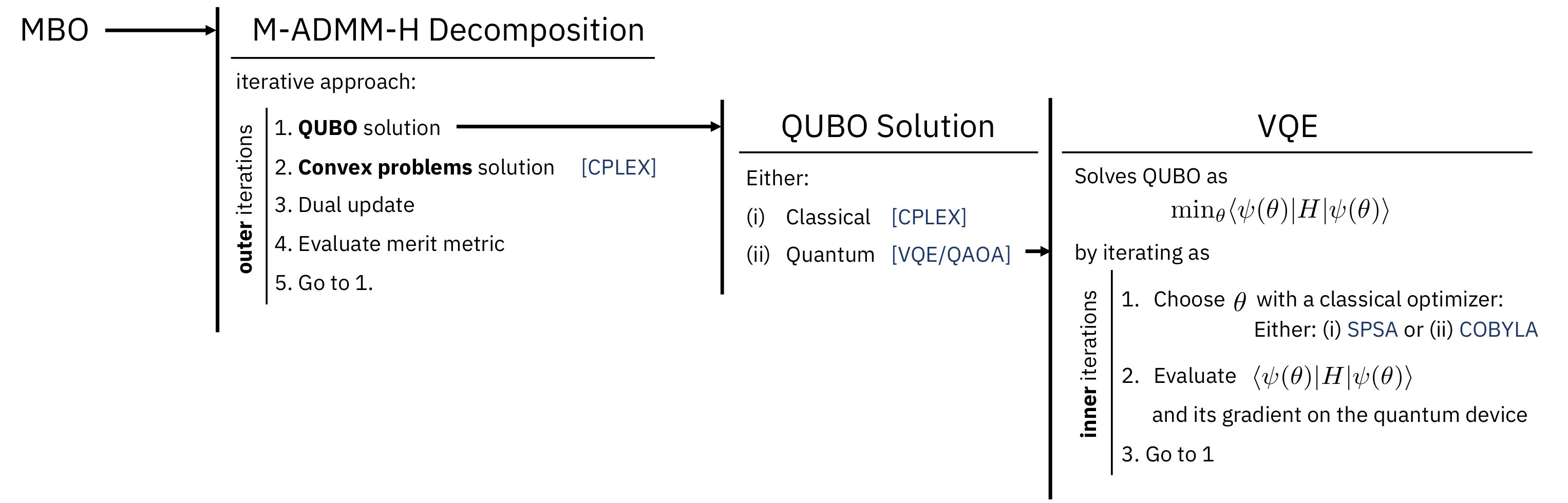}
\caption{Illustration diagram of \admm~with VQE as quantum solver. The optimization solvers adopted for the numerical results are specified (i.e., CPLEX, VQE, SPSA, and COBYLA). There are two nested loops for the selected implementation, specifically the outer ADMM loop with \revision{outer iterations}, and the inner VQE loop, \revision{with inner iterations}.}
\label{fig.appr}
\end{figure*}


The gap of the minimum-merit-value solution with value $v$ with respect to known optimal value $v^*$ is computed as $\frac{\lvert v - v^* \rvert}{1e-10+\lvert v^* \rvert}.$ In order to report the computational results, we have included: the number of \revision{binary decision variables (BinVars)},  
 the number of \revision{outer} iterations (IT) of \admm, the gap (Gap) to optimality, and percentage of \admm\ solution that are feasible (Feas) or optimal (Opt) with respect to the constraints and objective of the original constrained problem. \revision{For the simulations with VQE and QAOA, the number of binary decision variables corresponds to the number of qubits.}

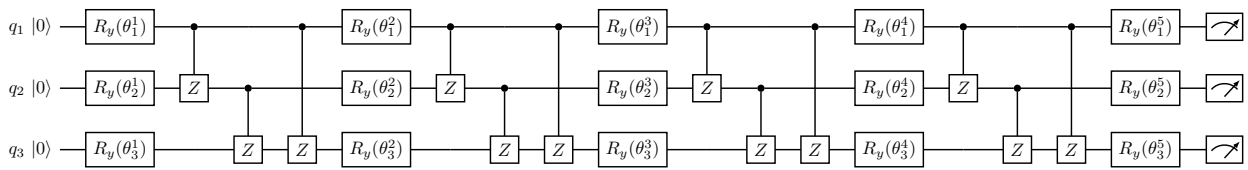
\begin{figure*}[t]
\centering
\begin{tikzpicture}
\node[scale=0.68] {
\begin{quantikz}
\lstick{$q_1 \, \ket{0}$} & 
\gate{R_y(\theta_1^1)} & \ctrl{1} & \qw & \ctrl{2} & 
\gate{R_y(\theta_1^2)} & \ctrl{1} & \qw & \ctrl{2} & 
\gate{R_y(\theta_1^3)} & \ctrl{1} & \qw & \ctrl{2} & 
\gate{R_y(\theta_1^4)} & \ctrl{1} & \qw & \ctrl{2} & 
\gate{R_y(\theta_1^5)} &
\meter{} \\
\lstick{$q_2 \, \ket{0}$} & 
\gate{R_y(\theta_2^1)} & \gate{Z}  & \ctrl{1} & \qw & 
\gate{R_y(\theta_2^2)} & \gate{Z}  & \ctrl{1} & \qw & 
\gate{R_y(\theta_2^3)} & \gate{Z}  & \ctrl{1} & \qw & 
\gate{R_y(\theta_2^4)} & \gate{Z}  & \ctrl{1} & \qw & 
\gate{R_y(\theta_2^5)} & 
\meter{} \\
\lstick{$q_3 \, \ket{0}$} & 
\gate{R_y(\theta_3^1)} & \qw & \gate{Z} & \gate{Z} & 
\gate{R_y(\theta_3^2)} & \qw & \gate{Z} & \gate{Z} & 
\gate{R_y(\theta_3^3)} & \qw & \gate{Z} & \gate{Z} & 
\gate{R_y(\theta_3^4)} & \qw & \gate{Z} & \gate{Z} & 
\gate{R_y(\theta_3^5)} & 
\meter{} \\
\end{quantikz}
};
\end{tikzpicture}
\caption{Prototype circuit used in the simulation results to evaluate $|\psi(\theta)\rangle = U(\theta) |0\rangle$, here exemplified for three qubits ($q = 3$) and a depth $d = 4$, consisting of $d+1$ layers. The first operations consists in single-qubit $Y$ rotations, with one variational parameter $\theta_{i}^j$ per qubit to determine the rotation angle. Each additional layer after the first contains entangling gates, more specifically controlled-$Z$ gates applied to all qubit pairs, followed by another set of single-qubit $Y$ rotations with one variational parameter each to represent the angle. The variational form is then parametrized over $q (d + 1)$ angles,  arranged in a vector $\theta = [\theta_i^j]_{i=1,\ldots,q; j = 1,\ldots,d+1}$.}
\label{fig.circuit}
\end{figure*}

\begin{remark}\label{rem:2}
We notice here that VQE and QAOA do not solve  (in general) a QUBO at optimality (and in this sense, they are not a perfect oracle), while CPLEX does (for the considered small instances). In addition, even in cases in which the quantum algorithm solves the QUBO at optimality, the optimizer may be different from CPLEX, since multiple equivalent solutions could exist. In general, then the solution of the quantum algorithm and CPLEX will be different when solving the same QUBO and the outer ADMM loop will be affected by it. In practice, using VQE or QAOA could either worsen or boost convergence: since \admm\ is in general a heuristic, small errors can be beneficial in some cases, while worsening performance in others.  

We notice that the choice of VQE and QAOA in this paper is due to the current technical status of quantum computing. In the future, better QUBO solvers may be available, e.g., based on (iterative) phase estimation, which might deliver optimal solutions at scale. 
\end{remark} 

\subsection{BP} \label{sec:bp}

We first discuss two implementation improvements to reduce computational complexity and foster convergence in the heuristic case for BP. 

\paragraph{Removing unnecessary decision variables.} Let $l$ be a lower bound on the number of bins required to pack all items (for example the continuous relaxation bound $\lceil \frac{\sum_{i,j} w_{ij}}{Q} \rceil$). Then, it is possible to discard variables $\chi_{1}, \dots, \chi_{l}$ from the mathematical formulation. In addition, it is not restrictive to assume $\xi_{1, 1} = 1$. With these observations, the number of decision variables required is $(mn-n) + (m-l)$. Typically, $n=m$, hence this boils down to $n^2-l$. The stronger the bound $l$ is, the fewer binary variables are introduced. In the current implementation, the continuous relaxation bound has been adopted.

\paragraph{Local search operator (LS).} 
To improve the convergence  of \admm~to solutions that are feasible for the equality constraints \eqref{BPP:capacity}, we have implemented a local search operator \cite{alvim1999local} to be applied to the solutions of the QUBO in the first block update of \eqref{algo:threeblock} and \eqref{algo:twoblock}. This operator is based on the Karmarkar-Karp Differencing Method \cite{michiels2003performance}, and it shuffles the assignment of items to pairs of bins in such a way to minimize the difference of the weights of the bin.

Bin Packing has been tested on \admm~on two groups of instances:

\begin{itemize}
	\item Small-sized: $n=2,3,4.$ Weights $w_j$ have been randomly picked in $[1, Q]$. The QUBO has been solved via VQE and CPLEX. 
	\item Scholl dataset \cite{scholl1997bison}, with $n=50$. \revision{We have considered $20$ instances of the dataset. For $10$ of the instances the bin capacity $Q$ is $100$, and the weights $w_j$ are sampled either from the interval $[1, 100]$, or the interval $[20, 100]$. For the remaining half of the instances, the weights are determined analogously, and the bin capacity is $120$.} 
\end{itemize}

On the Scholl dataset instances, the QUBO subproblem has been solved via CPLEX only, to evaluate the quality of \admm~solutions. The simulations on quantum devices are not of practical implementation at the moment, since the number of qubits in QUBO are $O(n^2)$ and would exceed the capabilities of current quantum technology.

\subsubsection{Small-sized dataset}

\paragraph{Simulations on classical devices}
For the simulations on CPLEX, \Cref{tab:bpp-toy-classical} reports the percentage of instances for which \admm~finds feasible or optimal solutions, grouped by the number of items of the instance. The 3-block \threeadmm~implementation is able to find feasible solutions for over $90\%$ of the instances. The search for optimal solutions becomes more difficult as the number of items increases, and for only $5\%$ of the $4$-items instances optimal solutions are found, and the gap to optimality is close to $70\%$ on the  $4$-items instances. For the two-block implementation \twoadmm~the increase of gap is less, however the search for feasible solutions is more difficult, as for  $63.33\%$ of the instances feasible solutions are found.

\begin{table*}[th!]
  \centering
  \sf
    \begin{tabular}{r|rrr|rrr}
          & \multicolumn{3}{c|}{\threeadmm}        & \multicolumn{3}{c}{\twoadmm}   \\\hline
    \multicolumn{1}{l|}{Items} & Gap & \multicolumn{1}{l}{Feas} & \multicolumn{1}{l|}{Opt} & Gap & \multicolumn{1}{l}{Feas} & \multicolumn{1}{l}{Opt} \\\hline
    2   & 0.00\%  &  100.00\%  &  100.00\%  & 25.00\% & 50.00\%  &  50.00\%  \\
3 & 15.83\% & 90.00\%  &  65.00\%  & 9.17\% & 90.00\%  &  75.00\%  \\
4 & 68.33\% &  100.00\%  &  5.00\%  & 20.42\% &  50.00\%  &  35.00\%  \\\hline
2, 3, 4  & 28.06\% &  95.08\%  &  55.74\% & 18.19\% &  63.33\%  &  53.33\%  \\
    \end{tabular}%
	\caption{Feasibility and optimality results of \admm\ on $60$ instances with $n=2,3,4$ and $Q=40$.}
\label{tab:bpp-toy-classical}
\end{table*}%


\paragraph{Simulations on quantum devices}
For the simulations in which QUBO is solved via VQE and QAOA, the classical solvers SPSA and COBYLA have been set with \revision{$10, 20, 50$ maximum inner iterations} on BP instances with $N=2, 3$ and $Q=40$. SPSA is known to be more computationally demanding than COBYLA, because it requires two function evaluations per iteration.  For each combination of values of $N$ and $Q$, $20$ instances have been generated with weights in $[1, Q]$, and average results for each group are reported for VQE in \Cref{tab:bpp-3-block-q}.
\rev{While the choice of $10$ maximum \revision{inner} iterations for the classical optimizer lowers the  computational time each ADMM \revision{outer} iteration, the convergence of ADMM is slowed down and the quality of the solution is also impacted negatively.}
  \revision{VQE-COBYLA} makes ADMM converge in $1$ \revision{outer} iteration to the optimal solution for instances with $2$ items with $20, 50$ maximum \revision{inner} iterations. Increasing the number of SPSA iterations is detrimental for the gap, feasibility and optimality of the instances: this is because SPSA runs for as many \revision{inner} iterations as the limit set in Qiskit. Invoking VQE with $50$ maximum \revision{inner} iterations in COBYLA, enables to increase by $40\%$ the number of instances with feasible solutions with $N=3$. Overall, the choice of SPSA as classical solver for VQE with $20$ \revision{inner} iterations is the best one in terms of solutions quality for these instances with $2$ and $3$ items, and outperforms the results obtained with CPLEX displayed in \Cref{tab:bpp-toy-classical}. This can be explained by the percentage of QUBO suproblems solved to optimality by VQE (column QUBO): while \revision{VQE-COBYLA} with $20$ or $50$ \revision{inner} iterations solves all QUBOs to optimality when $N=2$, \revision{VQE-SPSA} reports a non-optimal QUBO solution in a considerable percentage of the instances when $N=2$. \revision{It seems therefore beneficial for ADMM to solve a part of the QUBO suproblems in an inexact fashion}. For instances with $N=3$, the number of qubits increases and VQE hardly ever solves the QUBOs to optimality. \rev{Nevertheless, \threeadmm~converges to feasible and optimal solutions in all instances, with SPSA chosen as classical solver  (cf. Remark~\ref{rem:2}, and \Cref{sec:inexact}).} \rev{It is also interesting to note that \revision{VQE-SPSA} with $10$ SPSA \revision{inner} iterations solves the QUBO to optimality in $35\%$ more of the cases w.r.t. $20-50$ max \revision{inner} iterations, however this is detrimental to the gap, and optimality of the solutions.} 
The residuals are not guaranteed to decrease in each ADMM \revision{outer} iteration, as reported by \Cref{fig:conv-BP} on instance N3C40I8. In this case, \threeadmm~explores solutions with $3$ bins for about $70$ \revision{outer} iterations, and then converges to a non-optimal solution of lower value, which makes the residual equal to $0$.

\begin{figure}
	\centering
	\includegraphics[width = 0.48\textwidth]{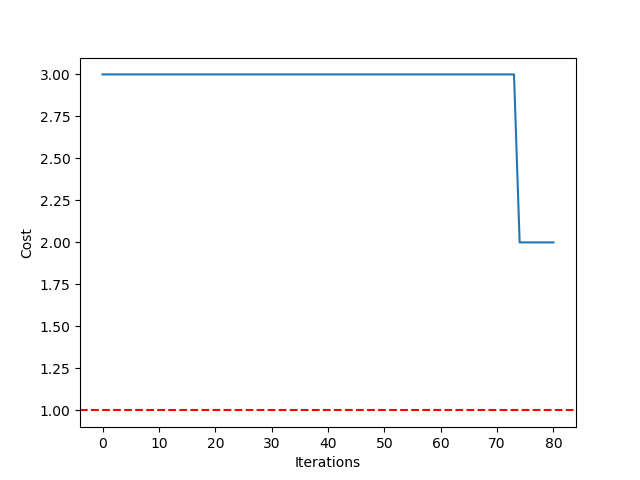}
	\includegraphics[width = 0.48\textwidth]{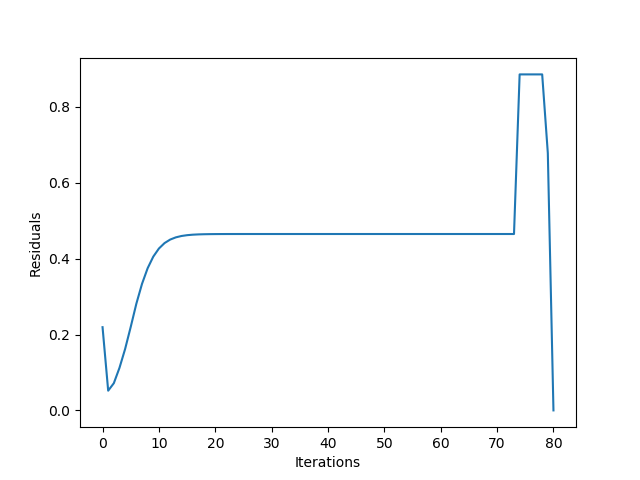}
	\caption{Plots of solution cost and residuals for instance N3C40I8, solved by \threeadmm~with COBYLA with $20$ \revision{inner} iterations. The optimal solution value is reported in red dashes.}
	\label{fig:conv-BP}
\end{figure}

\begin{table*}[th!]
  \centering
  \sf
  \footnotesize
    \begin{tabular}{lc|rrrrr|rrrrr}
          && \multicolumn{5}{c|}{SPSA}     & \multicolumn{5}{c}{COBYLA}   \\\hline
    Instance & \revision{BinVars} & \multicolumn{1}{l}{IT} & \multicolumn{1}{l}{Gap} & \multicolumn{1}{l}{Feas} & \multicolumn{1}{l}{Opt}  &  \multicolumn{1}{l|}{QUBO} & \multicolumn{1}{l}{IT} & \multicolumn{1}{l}{Gap} & \multicolumn{1}{l}{Feas} & \multicolumn{1}{l}{Opt} &  \multicolumn{1}{l}{QUBO} \\\hline
            \rev{N2C40IT10} & 2     & 1     & 50.00\% & 0.00\% & 0.00\% & 0.00\% & 9     & 0.00\% & 100.00\% & 100.00\% & 26.98\% \\
N2Q40IT20 & 2&  8  & 0.00\% & 100.00\% & 100.00\% & 41.81\% &  1     & 0.00\% & 100.00\% & 100.00\% & 100.00\%\\
N2Q40IT50  & 2&  13  & 35.00\% & 80.00\% & 65.00\%  & 64.87\% & 1     & 0.00\% & 100.00\% & 100.00\% & 100.00\%\\\hline
      \rev{N3C40IT10} & 7     & 115   & 21.67\% & 80.00\% & 70.00\% & 35.52\% & 69    & 30.00\% & 20.00\% & 10.00\% & 15.86\% \\
N3Q40IT20  & 7&  8  & 0.00\% & 100.00\% & 100.00\%  &  0.00\%  & 16 & 72.50\% & 80.00\% & 15.00\% & 0.26\% \\
N3Q40IT50 & 7& 6  & 14.17\% & 80.00\% & 75.00\%  & 0.87\% & 12  & 52.50\% & 80.00\% & 55.00\% & 0.00\% \\
    \end{tabular}%
    \caption{Average results of \threeadmm~on $40$ BP instances with $N=2,3$ and $Q=40$. The QUBO subproblems have been solved via VQE with SPSA and COBYLA solvers with \rev{$10$, $20$ and $50$} maximum \revision{inner} iterations.}
      \label{tab:bpp-3-block-q}%
\end{table*}%

\rev{
The results of \threeadmm~with QAOA as quantum solver are reported in Table \ref{tab:bpp-3-block-q-qaoa} in the Appendix. While the convergence of \threeadmm~is overall slower w.r.t. VQE, the quality of the solution obtained is similar when COBYLA performs the classical subroutines. 
}

On the same groups of BP instances, \twoadmm~has been tested, and average results are reported in \Cref{tab:bpp-2-block-q} for VQE. The convergence is overall slower than the $3$-block implementation in terms of number of \revision{outer} iterations, regardless of the classical solver called by VQE. SPSA makes \admm~obtain solutions with higher quality, when its maximum number of \revision{inner} iterations is set to $50$. COBYLA yields solutions with lower quality, in the case of $N=3$ and $50$ \revision{maximum inner iterations}. \rev{As observed for \threeadmm, the choice of $10$ maximum  inner iterations for SPSA and COBYLA delivers ADMM solutions with sensibly lower quality.} \rev{As observed for \threeadmm, a certain degree of inexactness in solving QUBOs is beneficial for the quality of the solutions delivered. In particular, for the instances with $3$ items, \twoadmm~delivers the best results when VQE solves $27\%$ of the QUBOs to optimality.}
\rev{
	The results of \twoadmm~with QAOA are shown in \Cref{tab:bpp-2-block-q-qaoa} in the Appendix. The choice of QAOA as quantum solver is beneficial in the simulations with COBYLA with $50$ inner iterations on instances with $3$ items: in this case, the best results in terms of gap, feasibility and optimality are obtained for \twoadmm.
}

\begin{table*}[th!]
  \centering
  \sf
  \footnotesize
    \begin{tabular}{lc|rrrrr|rrrrr}
          && \multicolumn{5}{c|}{SPSA}     & \multicolumn{5}{c}{COBYLA}   \\\hline
    Instance & BinVars & \multicolumn{1}{l}{IT} & \multicolumn{1}{l}{Gap} & \multicolumn{1}{l}{Feas} & \multicolumn{1}{l}{Opt}  &  \multicolumn{1}{l|}{QUBO} & \multicolumn{1}{l}{IT} & \multicolumn{1}{l}{Gap} & \multicolumn{1}{l}{Feas} & \multicolumn{1}{l}{Opt} &  \multicolumn{1}{l}{QUBO} \\\hline
       \rev{ N2C40IT10} & 2     & 1     & 50.00\% & 0.00\% & 0.00\% & 0.00\% & 6     & 5.00\% & 90.00\% & 90.00\% & 90.10\% \\
    N2Q40IT20 & 2 & 21    & 0.00\% & 100.00\% & 100.00\% & 86.15\% & 1     & 0.00\% & 100.00\% & 100.00\% & 100.00\% \\
    N2Q40IT50 & 2 & 15    & 0.00\% & 100.00\% & 100.00\% & 1.85\% & 1     & 0.00\% & 100.00\% & 100.00\% & 46.40\% \\\hline
        \rev{N3C40IT10} & 7     & 87    & 42.11\% & 63.16\% & 42.11\% & 11.99\% & 155   & 36.67\% & 30.00\% & 10.00\% & 10.00\% \\
    N3Q40IT20  & 7 & 14    & 29.17\% & 95.00\% & 65.00\% & 88.75\%  & 49    & 72.50\% & 80.00\% & 15.00\% & 100.00\% \\
    N3Q40IT50 & 7 & 5     & 7.50\% & 100.00\% & 90.00\% & 27.18\% & 11    & 41.67\% & 50.00\% & 50.00\% & 11.43\%\\   
    \end{tabular}%
\caption{Average results of \twoadmm~on $40$ BP instances with $N=2,3$ and $Q=40$. The QUBO subproblems have been solved via VQE with SPSA and COBYLA solvers with $10, 20$ and $50$ maximum inner iterations.}
  \label{tab:bpp-2-block-q}%
\end{table*}%

Finally, the VQE simulations where conducted on $3$ BP instances with $N=4, Q=4.$ In this case, \admm~cannot perform more than $2$ \revision{outer} iterations within the time limit of $1$ hour.
 We have also observed that, due to the size of the search state, VQE is not always able to find a solution where the equality constraints \eqref{BPP:assign} are satisfied. The number of \revision{inner} iterations of the classical solver invoked by VQE has to be set to a sufficiently large value that ensures to explore solutions without augmented Lagrangian penalty terms. As a representative example, \Cref{bpp-N4} displays the allocation of items to the bins on a BP instance (referred to as instance N$4$Q$4$) with weights $[2, 3, 2, 2],$ obtained from QUBO at outer iteration $1$ of \threeadmm. Since one of the items with weight $2$ is assigned twice in the solution obtained by \revision{VQE-SPSA} with $50$ \revision{inner} iterations, it is necessary to increase the \revision{inner} iterations to $100$ to obtain a solution where all items are assigned to one bin. In this case, the solution is feasible and optimal. 
 The time required to perform this outer \threeadmm~iteration goes from $2749.87
s$ in the $50$-\revision{inner}-iteration simulation case to $5380.40 s$ in the $100$-\revision{inner}-iterations case. This shows that the BP instances with $4$ items are computationally very demanding for the \threeadmm~algorithm.

\begin{figure}[h!]
	\begin{subfigure}{0.5\textwidth}
	\includegraphics[width=\textwidth]{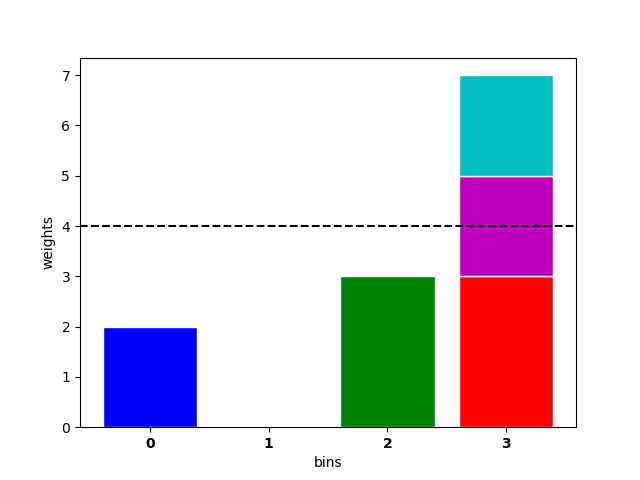}
	\label{bpp-N4-50}
\end{subfigure}
	\begin{subfigure}{0.5\textwidth}
			\includegraphics[width=\textwidth]{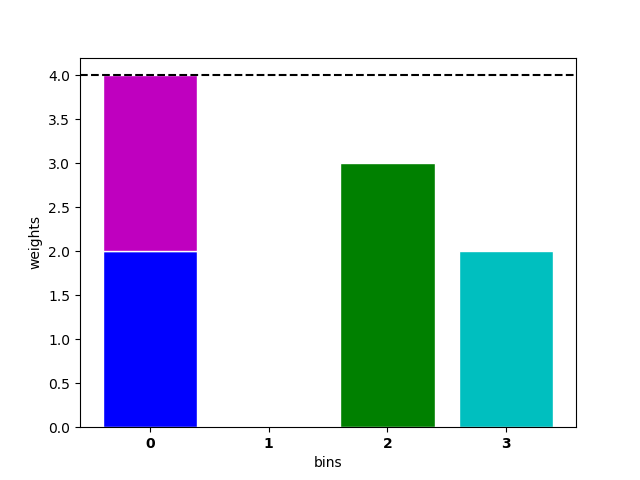}
					\label{bpp-N4-100}
		\end{subfigure}
				\caption{Solution found by VQE for the BP QUBO at outer iteration $1$ of \threeadmm, on instance N$4$Q$4$. The maximum number of SPSA \revision{inner} iterations is set to $50$ in the picture on the left, and to $100$ in the one on the right.}
			\label{bpp-N4}
\end{figure}


\subsubsection{Scholl dataset}

\Cref{tab:bpp-scholl} reports the results obtained with \admm, both without and with the local search (LS) operator described in \Cref{sec:bp}, with CPLEX as solver for QUBO. We have included  the percentage of time spent in solving the QUBO (column Block1), the convex subproblem (Block2), the convex and quadratic subproblem (Block3), and boolean indications for the feasibility (Feas) and optimality (Opt) of the \admm\ solution with respect to the constraints and objective of the original constrained problem. Without LS, \threeadmm\ takes on average $54$ \revision{outer} iterations to converge, and a feasible not optimal solution is found for all instances. The gap to optimality is $86.08\%$ and it is heavily depended on the capacity of the bins: the gap increases from $69.49\%$ on the instances with $Q=100$ to $102.68\%$ on the instances with $Q=120$. Applying LS on the QUBO solutions does not help to increase the solution quality, and in fact the same feasible solutions are obtained in the LS simulations. The advantage of LS in this case is to let \threeadmm\ converge in $1$ \revision{outer} iteration. The LS is instead extremely beneficial to find feasible solutions in the \twoadmm, and it enables to reach convergence within $21$ \revision{outer} iterations, on average. It is worthy to note that the $2$-block implementation enables to find solution with average gaps to optimality less than $50\%$ on those found by the $3$-block implementation, even if the convergence is often not reached in $500$ \revision{outer} iterations. In the $2$-block implementation, the gap is less dependent on the capacity of the bins. 


\begin{table*}[th!]
	\centering
	\sf
	\begin{tabular}{l|rrrr|rrrr}
		&\multicolumn{4}{c|}{No LS}      & \multicolumn{4}{|c}{With LS}   \\\hline
		Blocks &  \multicolumn{1}{l}{IT} & \multicolumn{1}{l}{Gap} & \multicolumn{1}{l}{Feas} & \multicolumn{1}{l|}{Opt} &  \multicolumn{1}{l}{IT} & \multicolumn{1}{l}{Gap} & \multicolumn{1}{l}{Feas} & \multicolumn{1}{l}{Opt} \\\hline
    3     & 54    & 86.08\% & 100.00\% & 0.00\% & 1     & 86.08\% & 100.00\% & 0.00\% \\
    2     & 471   & 30.17\% & 10.00\% & 0.00\% & 21    & 33.76\% & 100.00\% & 0.00\% \\
	\end{tabular}%
	\caption{Computational results of \admm\ on $20$ BP instances of the Scholl dataset \cite{scholl1997bison}. The QUBO subproblems are solved via CPLEX on the classical machine.}
	\label{tab:bpp-scholl}%
\end{table*}%

\subsection{MISK}

Mixed-Integer Setup Knapsack problem has been tested on \admm~on $2$ groups of instances. The first  group of instances, Group 1, has been generated by following the guidelines of \cite{altay2008exact}. To generate challenging MISK instances, the capacity utilization $\frac{\sum_{k=1}^K \sum_{t=1}^T D_{kt}}{P}$ is set to $2.5$, data correlation is medium (i. e., $D_{kt}  \in [1, 10], C_{kt} \in -[D_{kt}-2, D_{kt}+2 ] $), and setup costs $S_k$ are randomly sampled in $[40,60]$. A second group of instances, Group 2, has been generated with the aim to test \admm~in cases where the continuous decisions have an impact larger than the binary decisions on the solutions. To this end, the $S_k$  and values $C_{kt}$ have been lowered, specifically $S_k \in [0,1]$, and $C_{kt} \in [-60, -40]$. In both groups of instances, $T$ has been set to $10$, and the number of families $K$, corresponding to the number of qubits in the QUBO, ranges in the set $\{5, 8, 11, 14 \}.$\\
Both groups have been initially tested on \admm~with QUBO solved via CPLEX on a classical device. In this case, the feasible solution in which no item is assigned to the knapsack is very often the only feasible solution found, which can be arbitrarily far from the optimal value.
For the simulations with VQE, \Cref{tab:miskp3-g1} reports the average results obtained on $3$ instances for fixed $K$ in Group 1, with \threeadmm. While \threeadmm~with \revision{VQE-SPSA} fails to converge within $1$ hour for instances with $K \geq 8$, and $20$ SPSA \revision{inner} iterations, it converges with \revision{VQE-COBYLA} in a few \revision{outer} iterations, and produces more feasible solutions. However, \revision{VQE-SPSA} yields better results in terms of optimality gap, especially when the maximum number of SPSA \revision{inner} iterations is set to $20$. Feasible solutions are found for all instances with \revision{VQE-COBYLA} with $20$ COBYLA \revision{inner} iterations. \rev{The number of ADMM \revision{outer} iterations generally decreases with the increase of the \revision{inner} iterations of the classical optimizer.}

\begin{table*}[th!]
  \centering
  \sf
  \small
    \begin{tabular}{cl|rrrrr|rrrrr}
	& &  \multicolumn{5}{c|}{SPSA}    & \multicolumn{5}{c}{COBYLA} \\\hline
Instances  & BinVars &  \multicolumn{1}{c}{IT} & \multicolumn{1}{c}{Gap} & \multicolumn{1}{c}{Feas} & \multicolumn{1}{c}{Opt}  & \multicolumn{1}{c|}{QUBO}  &\multicolumn{1}{c}{IT} & \multicolumn{1}{c}{Gap} & \multicolumn{1}{c}{Feas} & \multicolumn{1}{c}{Opt} & \multicolumn{1}{c}{QUBO}  \\\hline
    \rev{K5IT10} & 5 & 19    & 62.13\% & 100.00\% & 33.33\% & 41.56\% & 6     & 100.00\% & 100.00\% & 0.00\% & 100.00\% \\
    K5IT20 &  5 & 11    & 76.20\% & 100.00\% & 0.00\%  & 25.87\% &  6    & 100.00\% & 100.00\% & 0.00\% & 100.00\%\\
    K5IT50 &  5 &  6     & 150.29\% & 100.00\% & 0.00\% & 52.80\% & 6    & 100.00\% & 100.00\% & 0.00\% & 100.00\% \\\hline
     \rev{K8IT10} & 8 &  22    & 82.41\% & 100.00\% & 0.00\% & 7.69\% & 6     & 546.00\% & 100.00\% & 0.00\% & 83.33\% \\
    K8IT20 &  8 &  27    & 38.74\% & 100.00\% & 0.00\%  & 8.03\% &  6     & 100.00\% & 100.00\% & 0.00\% & 100.00\% \\
    K8IT50 & 8 & 12    & 114.17\% & 100.00\% & 0.00\%  & 16.75\% & 6     & 100.00\% & 100.00\% & 0.00\% & 100.00\%  \\\hline
        \rev{K11IT10} & 11 & 63    & 42.80\% & 100.00\% & 0.00\% & 0.00\% & 87    & 96.39\% & 100.00\% & 0.00\% & 0.00\% \\
    K11IT20 & 11 &  12    & 98.57\% & 100.00\% & 0.00\% &  0.00\% & 6     & 100.00\% & 100.00\% & 0.00\% & 45.24\% \\
    K11IT50 & 11 & 5     & 94.66\% & 0.00\% & 0.00\%  & 2.90\% &  6     & 93.31\% & 100.00\% & 0.00\% & 18.18\% \\\hline
        \rev{K14IT10} & 14 & 31    & 39.55\% & 100.00\% & 0.00\% & 0.00\% & 8     & 66.83\% & 100.00\% & 0.00\% & 60.71\% \\
    K14IT20 & 14 & 6     & 80.91\% & 33.33\% & 0.00\% &  0.00\% & 6     & 100.00\% & 100.00\% & 0.00\% & 15.38\%  \\
    K14IT50 & 14 & 3     & 118.08\% & 0.00\% & 0.00\%  &  0.00\% & 6     & 170.25\% & 33.33\% & 0.00\% & 0.00\% \\      
  \end{tabular}%
  \caption{Computational results of \threeadmm\ on $12$ MISK instances in Group 1. The QUBO subproblems have been solved via VQE with SPSA and COBYLA solvers with \rev{$10$,} $20$ and $50$ maximum \revision{inner} iterations.}
  \label{tab:miskp3-g1}%
\end{table*}%

 The Group 2 instances are solved with average optimality gap of \rev{$18.11\%$} with \revision{VQE-SPSA}, as shown in \Cref{tab:miskp3-g2}, reporting a \rev{$65\%$} decrease of this metric with respect to Group 1. Hence, \threeadmm~finds solutions of higher quality in case the continuous decision variables play an important role in the MBO model. Using \revision{VQE-COBYLA} to solve the QUBO is beneficial for the larger-sized instances with $K=14$, since the average optimality gap drops to $12.09\%$ with $50$ maximum \revision{inner} iterations, in $1$ hour of computation.

Similarly to what observed for the BP problem, the percentage of QUBOs solved to optimality by VQE tends to decrease with the increase of the number of qubits. With \revision{VQE-COBYLA}, almost all QUBOs are solved to optimality for instances with up to $8$ qubits.  \rev{The exception is given by the simulation with $10$ \revision{inner} iterations for COBYLA and $K=8$. We observe that, for the Group 2 instances, a lower percentage of QUBOs solved to optimality corresponds to \admm~solutions with value closer to the optimal (cf. Remark \ref{rem:2}, and \Cref{sec:inexact})}. Almost half of the QUBOs are solved to optimality by \revision{VQE-COBYLA} on the instances with $11$ qubits, while \revision{VQE-SPSA} solves exactly less than $3\%$ of the QUBOs. \Cref{fig:conv-MISK} shows solution costs and value of the residuals reported in the \threeadmm~\revision{outer} iterations on instance K11T10I1 with $11$ qubits. The solution cost changes at each outer iteration in a non monotonic way, and \threeadmm~converges to a feasible solution in $10$ \revision{outer} iterations.

\begin{figure}[h!]
	\centering
	\includegraphics[width = 0.48\textwidth]{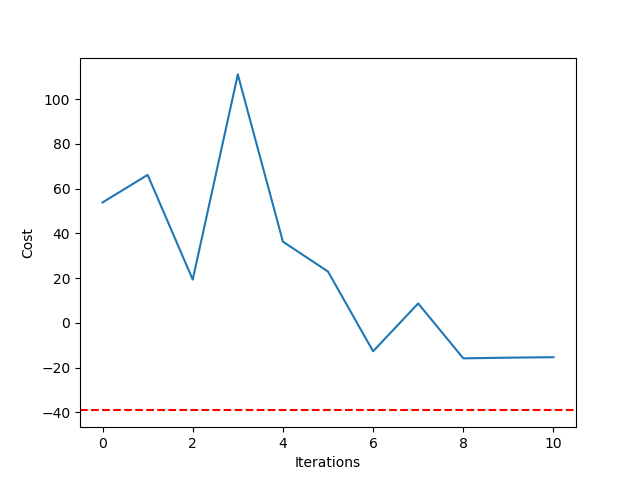}
	\includegraphics[width = 0.48\textwidth]{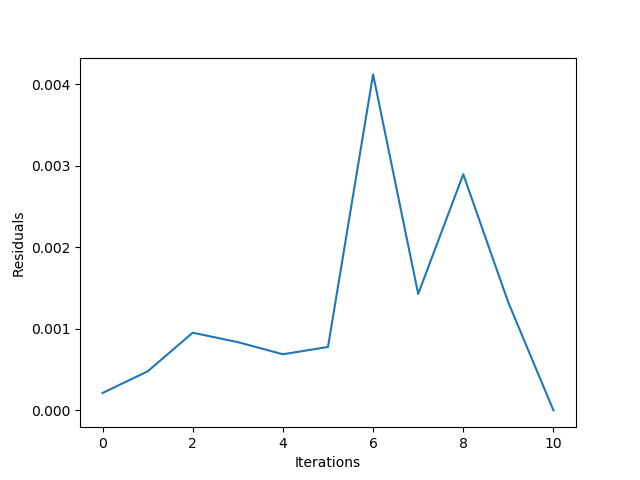}
	\caption{Plots of solution cost and residuals for instance K11T10I1, solved by \threeadmm~with COBYLA with $50$ inner iterations. The optimal solution value is reported in red dashes.}
	\label{fig:conv-MISK}
\end{figure}

\begin{table*}[th!]
	\centering
	\sf
	\small
	\begin{tabular}{lc|rrrrr|rrrrr}
	&& \multicolumn{5}{c|}{SPSA}    & \multicolumn{5}{c}{COBYLA} \\\hline
Instances  & BinVars & \multicolumn{1}{c}{IT} & \multicolumn{1}{c}{Gap} & \multicolumn{1}{c}{Feas} & \multicolumn{1}{c}{Opt} & \multicolumn{1}{c|}{QUBO} & \multicolumn{1}{c}{IT} & \multicolumn{1}{c}{Gap} & \multicolumn{1}{c}{Feas} & \multicolumn{1}{c}{Opt} & \multicolumn{1}{c}{QUBO}  \\\hline
 \rev{K5IT10} & 5 & 19    & 6.74\% & 100.00\% & 0.00\% & 23.70\% & 6     & 100.00\% & 100.00\% & 0.00\% & 100.00\% \\
    K5IT20 & 5 & 10    & 12.88\% & 100.00\% & 33.33\%  & 14.78\% &6     & 100.00\% & 100.00\% & 0.00\%  & 100.00\%\\
    K5IT50 & 5 & 7     & 47.30\% & 100.00\% & 0.00\% & 61.11\% & 6     & 100.00\% & 100.00\% & 0.00\% & 100.00\% \\\hline
        \rev{K8IT10} & 8 & 24    & 9.82\% & 100.00\% & 0.00\% & 12.06\% & 6     & 3.94\% & 100.00\% & 0.00\% & 83.33\% \\
    K8IT20  & 8 & 14    & 7.56\% & 100.00\% & 33.33\%  & 4.29\% & 6     & 100.00\% & 100.00\% & 0.00\% & 100.00\%\\
    K8IT50 & 8 & 11    & 20.54\% & 100.00\% & 0.00\%  &  14.88\% & 6     & 100.00\% & 100.00\% & 0.00\% & 100.00\%\\ \hline
        \rev{K11IT10} & 11 & 30    & 14.36\% & 66.67\% & 0.00\% & 0.00\% & 107   & 21.46\% & 100.00\% & 0.00\% & 0.00\% \\
    K11IT20  & 11 & 12    & 12.74\% & 100.00\% & 0.00\% & 0.00\% & 6     & 100.00\% & 100.00\% & 0.00\% & 47.62\%\\
    K11IT50 & 11 & 5     & 21.56\% & 0.00\% & 0.00\%  & 1.52\% & 8     & 22.42\% & 100.00\% & 0.00\% & 20.37\% \\ \hline
        \rev{K14IT10} & 14 & 19    & 18.19\% & 66.67\% & 0.00\% & 0.00\% & 7     & 28.72\% & 100.00\% & 0.00\% & 57.14\% \\
    K14IT20 & 14 & 6     & 21.57\% & 0.00\% & 0.00\% &0.00\% &6     & 100.00\% & 100.00\% & 0.00\% & 0.00\%\\
    K14IT50 & 14 &3     & 24.11\% & 0.00\% & 0.00\%  &0.00\% & 6     & 12.09\% & 0.00\% & 0.00\% & 0.00\% \\
    \end{tabular}%
  \caption{Computational results of \threeadmm\ on $12$ MISK instances in Group 2. The QUBO subproblems have been solved via VQE with SPSA and COBYLA solvers with \rev{$10$,} $20$ and $50$ maximum inner iterations.}
\label{tab:miskp3-g2}%
\end{table*}%



For the \twoadmm~implementation, the results are reported in \Cref{tab:miskp2-g1} and \Cref{tab:miskp2-g2}. As observed for \threeadmm, \twoadmm~with \revision{VQE-SPSA} delivers solutions with a lower average gap to optimality for the Group 2 instances. The convergence of \twoadmm~is slower than \threeadmm. Instances with $5$ and $11$ qubits and \revision{$20$ maximum COBYLA inner iterations} are solved by \twoadmm~within $1$ hour of computation. \rev{Setting $IT=10$ ensures faster simulations, at the price of solution quality.} Regarding the feasibility, on the one hand \twoadmm~with \revision{VQE-SPSA} and $IT \geq 20$ finds feasible solutions for $83.33\%$ of the Group 1 instances, and $91.67\%$ of the Group 2 instances, and on the other hand \twoadmm~with \revision{VQE-COBYLA} yields feasible solutions in respectively $75\%$ and $50\%$ of the cases. As in the \threeadmm~case, \revision{VQE-COBYLA} solves all QUBOs to optimality on instances with up to $8$ qubits\rev{, except for the case of $20$ maximum \revision{inner} iterations}. The percentage of QUBOs solved on the $11$-qubits instances increases substantially with respect to the \threeadmm~implementation.

\begin{table*}[th!]
  \centering
  \sf
  \small
	\begin{tabular}{lc|rrrrr|rrrrr}
	&& \multicolumn{5}{c|}{SPSA}    & \multicolumn{5}{c}{COBYLA} \\\hline
	Instances  & BinVars & \multicolumn{1}{c}{IT} & \multicolumn{1}{c}{Gap} & \multicolumn{1}{c}{Feas} & \multicolumn{1}{c}{Opt} & \multicolumn{1}{c|}{QUBO} & \multicolumn{1}{c}{IT} & \multicolumn{1}{c}{Gap} & \multicolumn{1}{c}{Feas} & \multicolumn{1}{c}{Opt} & \multicolumn{1}{c}{QUBO}  \\\hline
	     \rev{K5IT10} & 5 & 5     & 66.67\% & 100.00\% & 33.33\% & 70.00\% & 2     & 100.00\% & 100.00\% & 0.00\% & 100.00\% \\
    K5IT20 &  5 & 26    & 41.72\% & 100.00\% & 0.00\%  & 24.79\% & 22    & 100.00\% & 100.00\% & 0.00\% & 100.00\%
 \\
    K5IT50 &  5 & 1     & 205.00\% & 0.00\% & 0.00\%  & 72.22\%& 23    & 100.00\% & 100.00\% & 0.00\% & 100.00\%
 \\\hline
  \rev{K8IT10} & 8 & 9     & 104.15\% & 100.00\% & 0.00\% & 11.67\% & 2     & 546.00\% & 100.00\% & 0.00\% & 50.00\% \\
    K8IT20 & 8 &  63    & 0.01\% & 100.00\% & 0.00\%  & 11.82\% & 17    & 111.05\% & 100.00\% & 0.00\% & 100.00\%
 \\
    K8IT50 & 8 &  25    & 28.66\% & 100.00\% & 0.00\%  & 0.00\%
& 13    & 48.43\% & 0.00\% & 0.00\% & 100.00\%
 \\\hline
  \rev{K11IT10} & 11 & 37    & 75.31\% & 100.00\% & 0.00\% & 0.00\% & 93    & 96.39\% & 100.00\% & 0.00\% & 0.00\% \\
    K11IT20 & 11 &  12    & 22.94\% & 100.00\% & 0.00\%  & 0.00\%
& 22    & 100.00\% & 100.00\% & 0.00\% & 88.56\%
 \\
    K11IT50 & 11 &  8     & 77.39\% & 66.67\% & 0.00\% & 0.00\%
 & 23    & 100.00\% & 100.00\% & 0.00\% & 75.00\%
 \\\hline
  \rev{K14IT10} & 14 & 25    & 46.42\% & 100.00\% & 0.00\% & 0.00\% & 4     & 66.83\% & 100.00\% & 0.00\% & 25.00\% \\
    K14IT20 & 14 &  4     & 135.40\% & 33.33\% & 0.00\%  & 0.00\%
& 12    & 93.28\% & 0.00\% & 0.00\% & 0.00\%
\\
    K14IT50 & 14 &  3     & 132.06\% & 33.33\% & 0.00\% & 0.00\%
 & 6     & 93.77\% & 100.00\% & 0.00\%  & 0.00\% \\
    \end{tabular}%
  \caption{Computational results of \twoadmm\ on $12$ MISK instances in Group 1. The QUBO subproblems have been solved via VQE with SPSA and COBYLA solvers with \rev{$10$, }$20$ and $50$ maximum \revision{inner} iterations.}
  \label{tab:miskp2-g1}%
\end{table*}%

\begin{table*}[th!]
  \centering
  \sf
  \small
	\begin{tabular}{lc|rrrrr|rrrrr}
	&& \multicolumn{5}{c|}{SPSA}    & \multicolumn{5}{c}{COBYLA} \\\hline
	Instances  & BinVars & \multicolumn{1}{c}{IT} & \multicolumn{1}{c}{Gap} & \multicolumn{1}{c}{Feas} & \multicolumn{1}{c}{Opt} & \multicolumn{1}{c|}{QUBO} & \multicolumn{1}{c}{IT} & \multicolumn{1}{c}{Gap} & \multicolumn{1}{c}{Feas} & \multicolumn{1}{c}{Opt} & \multicolumn{1}{c}{QUBO}  \\\hline
	    \rev{K5IT10} & 5 & 2     & 100.00\% & 100.00\% & 0.00\% & 100.00\% & 2     & 100.00\% & 100.00\% & 0.00\% & 100.00\% \\
    K5IT20  & 5 & 84    & 0.03\% & 100.00\% & 0.00\%  & 13.24\% & 27    & 100.00\% & 100.00\% & 0.00\% & 	100.00\%
 \\
    K5IT50  & 5  & 86    & 0.00\% & 100.00\% & 0.00\%   & 64.60\% & 19    & 98.85\% & 66.67\% & 0.00\%  & 	100.00\%
\\\hline
    \rev{K8IT10} & 8 & 6     & 28.74\% & 100.00\% & 0.00\% & 2.78\% & 2     & 3.94\% & 100.00\% & 0.00\% & 50.00\% \\
    K8IT20  & 8  & 89    & 0.01\% & 100.00\% & 0.00\% & 0.00\%  & 2     & 22.32\% & 0.00\% & 0.00\% & 	100.00\%
 \\
    K8IT50  & 8  & 29    & 10.53\% & 100.00\% & 0.00\% & 8.33\%  & 1     & 16.86\% & 0.00\% & 0.00\% & 	100.00\%
 \\\hline
     \rev{K11IT10} & 11& 8     & 31.70\% & 66.67\% & 0.00\% & 0.00\% & 113   & 13.54\% & 100.00\% & 0.00\% & 0.00\% \\
    K11IT20 &  11 &  36    & 6.38\% & 100.00\% & 0.00\% & 0.00\%  & 27    & 100.00\% & 100.00\% & 0.00\% & 	91.83\%
 \\
    K11IT50 &  11 & 12    & 4.78\% & 100.00\% & 0.00\% & 0.00\%   & 27    & 100.00\% & 100.00\% & 0.00\%  & 	75.00\%
\\\hline
    \rev{K14IT10} & 14 & 20    & 11.63\% & 66.67\% & 0.00\% & 0.00\% & 4     & 28.72\% & 100.00\% & 0.00\% & 25.00\% \\
    K14IT20 &  14 & 5     & 14.13\% & 66.67\% & 0.00\% & 0.00\%   & 9     & 37.42\% & 0.00\% & 0.00\%  & 	0.00\%
\\
    K14IT50 &  14 & 3     & 26.63\% & 0.00\% & 0.00\% & 0.00\%  & 6     & 27.52\% & 33.33\% & 0.00\%  & 	0.00\%
\\
    \end{tabular}%
  \caption{Computational results of \twoadmm\ on $12$ MISK instances in Group 2. The QUBO subproblems have been solved via VQE with SPSA and COBYLA solvers with \rev{$10$, }$20$ and $50$ maximum \revision{inner} iterations.}
\label{tab:miskp2-g2}%
\end{table*}%

\rev{
The simulations  of \admm~with QAOA as quantum solver are reported in Tables \ref{tab:miskp3-g1-QAOA}, \ref{tab:miskp3-g2-QAOA}, \ref{tab:miskp2-g1-QAOA}, \ref{tab:miskp2-g2-QAOA} in the Appendix. The results further corroborate the claim that for Group 2 instances, \admm~find solutions of higher quality. This is observed in the drop of the solution gap in the \revision{QAOA-SPSA} simulations, between Group 1 and Group 2 instances. The gap drop corresponds to $76\%$ for \threeadmm, and $51\%$ for \twoadmm. Hence, the impact  of continuous decision variables in the convergence of \admm~could deserve more future studies.
}

\section{Conclusions}\label{sec:concl}

In this work, we have proposed an iterative heuristic method \admm, based on Alternating Direction Method of Multipliers, to solve MBOs on current noisy quantum devices as well as on classical computers whenever a QUBO solver is available. The method relies on a decomposition of MBO into a QUBO subproblem, which can be tackled via quantum optimization solvers such as VQE and QAOA, and convex subproblems. This enables to extend the range of mathematical optimization problems that can be solved on quantum devices.  The method has been tested via the Qiskit framework with VQE as quantum QUBO solver on two representative MBO problems, namely Bin Packing Problem, and Mixed-Integer Setup Knapsack Problem. The simulations indicated the effectiveness of \admm~ in finding solutions feasible for the MBO formulations. In particular, for Bin Packing instances with $2$ and $3$ items, feasible solutions are found with an average optimality gap of at most $7.50\%$. In this case, setting SPSA in \twoadmm~as the VQE solver with $50$ iterations delivers the best results. On MISK instances, VQE is beneficial to explore feasible solutions different to a trivial one found via the classical computation with CPLEX. \rev{It has also been highlighted that \threeadmm~finds solutions of higher quality in case the continuous decision variables play an important role in the MBO model, and this could deserve future investigation.}

 It is important to observe that \admm~is a heuristic optimization algorithm for a class of MBO formulations, and it is not tailored to the two applications addressed in this paper, namely BP and MISK problems; therefore the results in terms of feasibility are not trivial on these combinatorial problems.

In theory, we have presented formal requirements under which \threeadmm~is guaranteed to converge to a stationary point of a pertinent augmented Lagrangian, which applies on quantum and classical computers alike. In practice, we have offer a glimpse on current research in combinatorial optimization in quantum computing, along with assumptions, challenges, and open problems.

Future works can include the investigation of \rev{the impact of continuous decision variables in the \admm~convergence,} the integration of techniques to enforce the feasibility of equality constraints of MBO in the QUBO subproblems \cite{wang2019xy}, different decomposition approaches to devise QUBO subproblems, the combination of ADMM with slack variable approaches~\cite{braine2019quantum}, and alternative decomposition approaches to devise QUBO subproblems for MBO.

\section*{Acknowledgements}
The authors are grateful to Jakub Marecek, Martin Mevissen, and Stefan Woerner  at IBM Quantum, which provided constructive feedback on the manuscript.

\bibliographystyle{plain}

\bibliography{admm}

\onecolumn\newpage
\appendix
\section{QAOA simulations}\label{sec:app-qaoa}

\label{sec:app-qaoa}

\begin{table*}[th!]
	\centering
	\sf
	\footnotesize
	\begin{tabular}{lc|rrrrr|rrrrr}
		&& \multicolumn{5}{c|}{SPSA}     & \multicolumn{5}{c}{COBYLA}   \\\hline
		Instance & BinVars & \multicolumn{1}{l}{IT} & \multicolumn{1}{l}{Gap} & \multicolumn{1}{l}{Feas} & \multicolumn{1}{l}{Opt}  &  \multicolumn{1}{l|}{QUBO} & \multicolumn{1}{l}{IT} & \multicolumn{1}{l}{Gap} & \multicolumn{1}{l}{Feas} & \multicolumn{1}{l}{Opt} &  \multicolumn{1}{l}{QUBO} \\\hline
		N2C40IT10 & 2     & 1     & 50.00\% & 0.00\% & 0.00\% & 0.00\% & 10    & 0.00\% & 100.00\% & 100.00\% & 93.15\% \\
		N2C40IT20 & 2     & 273   & 0.00\% & 100.00\% & 100.00\% & 48.26\% & 1     & 0.00\% & 100.00\% & 100.00\% & 100.00\% \\
		N2C40IT50 & 2     & 137   & 0.00\% & 100.00\% & 100.00\% & 70.05\% & 1     & 0.00\% & 100.00\% & 100.00\% & 100.00\% \\\hline
		N3C40IT10 & 7     & 110   & 40.35\% & 63.16\% & 52.63\% & 3.06\% & 101   & 36.67\% & 20.00\% & 10.00\% & 0.00\% \\
		N3C40IT20 & 7     & 55    & 17.54\% & 89.47\% & 78.95\% & 0.00\% & 51    & 67.50\% & 60.00\% & 15.00\% & 0.26\% \\
		N3C40IT50 & 7     & 20    & 24.56\% & 57.89\% & 57.89\% & 1.03\% & 57    & 21.67\% & 80.00\% & 60.00\% & 0.14\% \\
	\end{tabular}%
	\caption{\rev{Average results of \threeadmm~on $40$ BP instances with $N=2,3$ and $Q=40$. The QUBO subproblems have been solved via QAOA with SPSA and COBYLA solvers with \rev{$10$, $20$ and $50$} maximum iterations.}}
	\label{tab:bpp-3-block-q-qaoa}%
\end{table*}%

\begin{table*}[th!]
	\centering
	\sf
	\footnotesize
	\begin{tabular}{lc|rrrrr|rrrrr}
		&& \multicolumn{5}{c|}{SPSA}     & \multicolumn{5}{c}{COBYLA}   \\\hline
		Instance & BinVars & \multicolumn{1}{l}{IT} & \multicolumn{1}{l}{Gap} & \multicolumn{1}{l}{Feas} & \multicolumn{1}{l}{Opt}  &  \multicolumn{1}{l|}{QUBO} & \multicolumn{1}{l}{IT} & \multicolumn{1}{l}{Gap} & \multicolumn{1}{l}{Feas} & \multicolumn{1}{l}{Opt} &  \multicolumn{1}{l}{QUBO} \\\hline
		N2C40IT10 & 2     & 1     & 50.00\% & 0.00\% & 0.00\% & 0.00\% & 7     & 0.00\% & 100.00\% & 100.00\% & 88.39\% \\
		N2C40IT20 & 2     & 500   & 0.00\% & 100.00\% & 100.00\% & 62.99\% & 1     & 0.00\% & 100.00\% & 100.00\% & 100.00\% \\
		N2C40IT50 & 2     & 52    & 0.00\% & 100.00\% & 100.00\% & 69.67\% & 1     & 0.00\% & 100.00\% & 100.00\% & 100.00\% \\\hline
		N3C40IT10 & 7     & 86    & 47.37\% & 63.16\% & 31.58\% & 6.87\% & 106   & 41.67\% & 20.00\% & 10.00\% & 10.00\% \\
		N3C40IT20 & 7     & 59    & 12.28\% & 84.21\% & 73.68\% & 24.42\% & 75    & 62.50\% & 50.00\% & 15.00\% & 10.46\% \\
		N3C40IT50 & 7     & 21    & 25.44\% & 68.42\% & 68.42\% & 17.56\% & 83    & 26.67\% & 90.00\% & 65.00\% & 20.69\% \\  
	\end{tabular}%
	\caption{\rev{Average results of \twoadmm~on $40$ BP instances with $N=2,3$ and $Q=40$. The QUBO subproblems have been solved via QAOA with SPSA and COBYLA solvers with $10, 20$ and $50$ maximum iterations.}}
	\label{tab:bpp-2-block-q-qaoa}%
\end{table*}%

\begin{table*}[th!]
	\centering
	\sf
	\footnotesize
	\begin{tabular}{cl|rrrrr|rrrrr}
		& &  \multicolumn{5}{c|}{SPSA}    & \multicolumn{5}{c}{COBYLA} \\\hline
		Instances  & BinVars &  \multicolumn{1}{c}{IT} & \multicolumn{1}{c}{Gap} & \multicolumn{1}{c}{Feas} & \multicolumn{1}{c}{Opt}  & \multicolumn{1}{c|}{QUBO}  &\multicolumn{1}{c}{IT} & \multicolumn{1}{c}{Gap} & \multicolumn{1}{c}{Feas} & \multicolumn{1}{c}{Opt} & \multicolumn{1}{c}{QUBO}  \\\hline
    K5IT10 & 5     & 19    & 62.13\% & 100.00\% & 33.33\% & 41.56\% & 6     & 100.00\% & 100.00\% & 0.00\% & 100.00\% \\
    K5IT20 & 5     & 11    & 4.91\% & 100.00\% & 33.33\% & 25.87\% & 6     & 100.00\% & 100.00\% & 0.00\% & 100.00\% \\
    K5IT50 & 5     & 7     & 198.19\% & 100.00\% & 0.00\% & 52.80\% & 6     & 100.00\% & 100.00\% & 0.00\% & 100.00\% \\\hline
    K8IT10 & 8     & 14    & 103.15\% & 100.00\% & 0.00\% & 7.69\% & 6     & 546.00\% & 100.00\% & 0.00\% & 100.00\% \\
    K8IT20 & 8     & 26    & 55.69\% & 100.00\% & 0.00\% & 7.55\% & 6     & 100.00\% & 100.00\% & 0.00\% & 100.00\% \\
    K8IT50 & 8     & 11    & 136.61\% & 66.67\% & 0.00\% & 45.42\% & 6     & 100.00\% & 100.00\% & 0.00\% & 100.00\% \\\hline
    K11IT10 & 11    & 58    & 33.30\% & 100.00\% & 0.00\% & 0.00\% & 7     & 118.73\% & 100.00\% & 0.00\% & 100.00\% \\
    K11IT20 & 11    & 34    & 116.66\% & 100.00\% & 0.00\% & 0.00\% & 8     & 117.00\% & 100.00\% & 0.00\% & 34.52\% \\
    K11IT50 & 11    & 14    & 80.41\% & 66.67\% & 0.00\% & 10.19\% & 11    & 81.00\% & 100.00\% & 0.00\% & 18.18\% \\\hline
    K14IT10 & 14    & 30    & 48.90\% & 100.00\% & 0.00\% & 0.00\% & 8     & 66.83\% & 100.00\% & 0.00\% & 100.00\% \\
    K14IT20 & 14    & 12    & 73.31\% & 66.67\% & 0.00\% & 0.00\% & 41    & 54.23\% & 100.00\% & 0.00\% & 0.00\% \\
    K14IT50 & 14    & 8     & 116.72\% & 66.67\% & 0.00\% & 0.00\% & 19    & 51.46\% & 100.00\% & 0.00\% & 0.00\% \\
	\end{tabular}%
	\caption{\rev{Computational results of \threeadmm\ on $12$ MISK instances in Group 1. The QUBO subproblems have been solved via QAOA with SPSA and COBYLA solvers with $10$, $20$ and $50$ maximum iterations.}}
	\label{tab:miskp3-g1-QAOA}%
\end{table*}%

\begin{table*}[th!]
	\centering
	\sf
	\footnotesize
	\begin{tabular}{cl|rrrrr|rrrrr}
		& &  \multicolumn{5}{c|}{SPSA}    & \multicolumn{5}{c}{COBYLA} \\\hline
		Instances  & BinVars &  \multicolumn{1}{c}{IT} & \multicolumn{1}{c}{Gap} & \multicolumn{1}{c}{Feas} & \multicolumn{1}{c}{Opt}  & \multicolumn{1}{c|}{QUBO}  &\multicolumn{1}{c}{IT} & \multicolumn{1}{c}{Gap} & \multicolumn{1}{c}{Feas} & \multicolumn{1}{c}{Opt} & \multicolumn{1}{c}{QUBO}  \\\hline
		K5IT10 & 5     & 19    & 6.74\% & 100.00\% & 0.00\% & 23.70\% & 6     & 100.00\% & 100.00\% & 0.00\% & 100.00\% \\
		K5IT20 & 5     & 24    & 9.80\% & 100.00\% & 0.00\% & 14.78\% & 6     & 100.00\% & 100.00\% & 0.00\% & 100.00\% \\
		K5IT50 & 5     & 7     & 30.65\% & 100.00\% & 0.00\% & 61.11\% & 6     & 100.00\% & 100.00\% & 0.00\% & 100.00\% \\\hline
		K8IT10 & 8     & 24    & 9.82\% & 100.00\% & 0.00\% & 12.06\% & 6     & 3.94\% & 100.00\% & 0.00\% & 100.00\% \\
		K8IT20 & 8     & 49    & 2.07\% & 100.00\% & 0.00\% & 3.41\% & 6     & 100.00\% & 100.00\% & 0.00\% & 100.00\% \\
		K8IT50 & 8     & 21    & 12.53\% & 100.00\% & 0.00\% & 14.88\% & 6     & 100.00\% & 100.00\% & 0.00\% & 100.00\% \\\hline
		K11IT10 & 11    & 47    & 8.59\% & 100.00\% & 0.00\% & 0.00\% & 10    & 13.54\% & 33.33\% & 0.00\% & 0.00\% \\
		K11IT20 & 11    & 45    & 3.22\% & 100.00\% & 0.00\% & 0.00\% & 7     & 29.84\% & 100.00\% & 0.00\% & 19.84\% \\
		K11IT50 & 11    & 13    & 14.58\% & 100.00\% & 0.00\% & 0.00\% & 10    & 16.11\% & 100.00\% & 0.00\% & 20.37\% \\\hline
		K14IT10 & 14    & 30    & 10.47\% & 100.00\% & 0.00\% & 0.00\% & 5     & 28.72\% & 100.00\% & 0.00\% & 35.00\% \\
		K14IT20 & 14    & 18    & 12.92\% & 100.00\% & 0.00\% & 0.00\% & 41    & 9.56\% & 100.00\% & 0.00\% & 0.00\% \\
		K14IT50 & 14    & 6     & 21.88\% & 33.33\% & 0.00\% & 0.00\% & 20    & 9.63\% & 100.00\% & 0.00\% & 0.00\% \\
	\end{tabular}%
	\caption{\rev{Computational results of \threeadmm\ on $12$ MISK instances in Group 2. The QUBO subproblems have been solved via QAOA with SPSA and COBYLA solvers with $10$, $20$ and $50$ maximum iterations.}}
	\label{tab:miskp3-g2-QAOA}%
\end{table*}%

\begin{table*}[th!]
	\centering
	\sf
	\footnotesize
	\begin{tabular}{cl|rrrrr|rrrrr}
		& &  \multicolumn{5}{c|}{SPSA}    & \multicolumn{5}{c}{COBYLA} \\\hline
		Instances  & BinVars &  \multicolumn{1}{c}{IT} & \multicolumn{1}{c}{Gap} & \multicolumn{1}{c}{Feas} & \multicolumn{1}{c}{Opt}  & \multicolumn{1}{c|}{QUBO}  &\multicolumn{1}{c}{IT} & \multicolumn{1}{c}{Gap} & \multicolumn{1}{c}{Feas} & \multicolumn{1}{c}{Opt} & \multicolumn{1}{c}{QUBO}  \\\hline		
    K5IT10 & 5     & 5     & 66.67\% & 100.00\% & 33.33\% & 70.00\% & 2     & 100.00\% & 100.00\% & 0.00\% & 100.00\% \\
    K5IT20 & 5     & 2     & 59.92\% & 100.00\% & 0.00\% & 61.11\% & 2     & 100.00\% & 100.00\% & 0.00\% & 100.00\% \\
    K5IT50 & 5     & 3     & 40.25\% & 100.00\% & 33.33\% & 41.67\% & 2     & 100.00\% & 100.00\% & 0.00\% & 100.00\% \\\hline
    K8IT10 & 8     & 8     & 123.85\% & 100.00\% & 0.00\% & 13.06\% & 2     & 546.00\% & 100.00\% & 0.00\% & 100.00\% \\
    K8IT20 & 8     & 29    & 80.86\% & 100.00\% & 0.00\% & 16.67\% & 2     & 100.00\% & 100.00\% & 0.00\% & 100.00\% \\
    K8IT50 & 8     & 2     & 78.82\% & 100.00\% & 0.00\% & 16.67\% & 2     & 100.00\% & 100.00\% & 0.00\% & 100.00\% \\\hline
    K11IT10 & 11    & 49    & 68.36\% & 100.00\% & 0.00\% & 0.00\% & 66    & 121.62\% & 100.00\% & 0.00\% & 100.00\% \\
    K11IT20 & 11    & 12    & 64.91\% & 100.00\% & 0.00\% & 0.00\% & 5     & 121.62\% & 100.00\% & 0.00\% & 24.44\% \\
    K11IT50 & 11    & 17    & 101.24\% & 100.00\% & 0.00\% & 1.96\% & 35    & 110.19\% & 100.00\% & 0.00\% & 0.00\% \\\hline
    K14IT10 & 14    & 11    & 110.14\% & 33.33\% & 0.00\% & 0.00\% & 4     & 66.83\% & 100.00\% & 0.00\% & 100.00\% \\
    K14IT20 & 14    & 6     & 122.53\% & 33.33\% & 0.00\% & 0.00\% & 1     & 71.15\% & 0.00\% & 0.00\% & 0.00\% \\
    K14IT50 & 14    & 5     & 44.43\% & 66.67\% & 0.00\% & 0.00\% & 19    & 46.81\% & 100.00\% & 0.00\% & 0.00\% \\
	\end{tabular}%
	\caption{\rev{Computational results of \twoadmm\ on $12$ MISK instances in Group 1. The QUBO subproblems have been solved via QAOA with SPSA and COBYLA solvers with $10$, $20$ and $50$ maximum iterations.}}
	\label{tab:miskp2-g1-QAOA}%
\end{table*}%

\begin{table*}[th!]
	\centering
	\sf
	\footnotesize
	\begin{tabular}{cl|rrrrr|rrrrr}
		& &  \multicolumn{5}{c|}{SPSA}    & \multicolumn{5}{c}{COBYLA} \\\hline
		Instances  & BinVars &  \multicolumn{1}{c}{IT} & \multicolumn{1}{c}{Gap} & \multicolumn{1}{c}{Feas} & \multicolumn{1}{c}{Opt}  & \multicolumn{1}{c|}{QUBO}  &\multicolumn{1}{c}{IT} & \multicolumn{1}{c}{Gap} & \multicolumn{1}{c}{Feas} & \multicolumn{1}{c}{Opt} & \multicolumn{1}{c}{QUBO}  \\\hline
K5IT10 & 5     & 2     & 100.00\% & 100.00\% & 0.00\% & 100.00\% & 2     & 100.00\% & 100.00\% & 0.00\% & 100.00\% \\
K5IT20 & 5     & 4     & 57.44\% & 100.00\% & 0.00\% & 31.94\% & 2     & 100.00\% & 100.00\% & 0.00\% & 100.00\% \\
K5IT50 & 5     & 2     & 67.05\% & 100.00\% & 0.00\% & 50.00\% & 2     & 100.00\% & 100.00\% & 0.00\% & 100.00\% \\\hline
K8IT10 & 8     & 6     & 28.74\% & 100.00\% & 0.00\% & 2.78\% & 2     & 3.94\% & 100.00\% & 0.00\% & 100.00\% \\
K8IT20 & 8     & 6     & 38.58\% & 100.00\% & 0.00\% & 16.67\% & 2     & 100.00\% & 100.00\% & 0.00\% & 100.00\% \\
K8IT50 & 8     & 2     & 49.56\% & 100.00\% & 0.00\% & 33.33\% & 2     & 100.00\% & 100.00\% & 0.00\% & 100.00\% \\\hline
K11IT10 & 11    & 47    & 8.59\% & 100.00\% & 0.00\% & 0.00\% & 10    & 13.54\% & 33.33\% & 0.00\% & 0.00\% \\
K11IT20 & 11    & 26    & 9.60\% & 100.00\% & 0.00\% & 3.04\% & 17    & 14.23\% & 33.33\% & 0.00\% & 0.00\% \\
K11IT50 & 11    & 14    & 15.56\% & 100.00\% & 0.00\% & 6.67\% & 23    & 13.90\% & 100.00\% & 0.00\% & 6.06\% \\\hline
K14IT10 & 14    & 30    & 10.47\% & 100.00\% & 0.00\% & 0.00\% & 5     & 28.72\% & 100.00\% & 0.00\% & 35.00\% \\
K14IT20 & 14    & 16    & 18.54\% & 100.00\% & 0.00\% & 0.00\% & 2     & 52.21\% & 100.00\% & 0.00\% & 0.00\% \\
K14IT50 & 14    & 8     & 24.77\% & 100.00\% & 0.00\% & 0.00\% & 17    & 9.45\% & 100.00\% & 0.00\% & 0.00\% \\		
	\end{tabular}%
	\caption{\rev{Computational results of \twoadmm\ on $12$ MISK instances in Group 2. The QUBO subproblems have been solved via QAOA with SPSA and COBYLA solvers with $10$, $20$ and $50$ maximum iterations.}}
	\label{tab:miskp2-g2-QAOA}%
\end{table*}%
\end{document}